\documentclass[twocolumn,showpacs,nofootinbib,superscriptaddress]{revtex4-1}

\usepackage[nointlimits]{amsmath}
\usepackage{hyperref}
\usepackage{amsfonts}
\usepackage{amssymb}
\usepackage{amsmath}
\usepackage{amsthm}
\usepackage{latexsym}
\usepackage{graphicx}
\usepackage{float}
\usepackage{color}
\usepackage{layout}
\usepackage[english]{babel}

\usepackage{wrapfig}

\def\LL{\mathcal{L}}
\def\x{\mathbf{x}}
\def\y{\mathbf{y}}

\def\V{\mathcal{V}}
\def\Y{\mathcal{Y}}
\def\F{\mathcal{F}}
\def\G{\mathcal{G}}
\def\eps{\varepsilon}

\begin{document}

\title{Constraining a nonminimally coupled curvature-matter gravity model with ocean experiments}

\author{Riccardo March}
\email{r.march@iac.cnr.it}
\affiliation{Istituto per le Applicazioni del Calcolo, CNR, Via dei Taurini 19, 00185 Roma, Italy}
\affiliation{INFN - Laboratori Nazionali di Frascati (LNF), Via E. Fermi 40, Frascati 00044 Roma, Italy}

\author{Orfeu Bertolami}
\email{orfeu.bertolami@fc.up.pt}
\affiliation{Departamento de F\'isica e Astronomia and Centro de F\'isica do Porto,\\Faculdade de Ci\^encias da Universidade do Porto, Rua do Campo Alegre 687, 4169-007 , Porto, Portugal}

\author{Marco Muccino}
\email{Marco.Muccino@lnf.infn.it}
\affiliation{INFN - Laboratori Nazionali di Frascati (LNF), Via E. Fermi 40, Frascati 00044 Roma, Italy}

\author{Rodrigo Baptista}
\email{rodrigo.baptista@fc.up.pt}
\affiliation{Departamento de F\'isica e Astronomia and Centro de F\'isica do Porto,\\Faculdade de Ci\^encias da Universidade do Porto, Rua do Campo Alegre 687, 4169-007 , Porto, Portugal}

\author{Simone Dell'Agnello}
\email{simone.dellagnello@lnf.infn.it}
\affiliation{INFN - Laboratori Nazionali di Frascati (LNF), Via E. Fermi 40, Frascati 00044 Roma, Italy}

\date{today}

\begin{abstract}
We examine the constraints on the Yukawa regime from the nonminimally coupled curvature-matter gravity theory arising from deep underwater ocean experiments.
We consider the geophysical experiment of Zumberge et al. of 1991 \cite{Zum} for searching deviations of Newton's inverse square law in ocean. 
In the context of nonminimally coupled curvature-matter theory of gravity the results of Zumberge et al. can be used to obtain an upper bound both
on the strength $\alpha$ and range $\lambda$ of the Yukawa potential arising from the nonrelativistic limit of the nonminimally coupled theory.
The existence of an upper bound on $\lambda$ is related to the presence of an extra force, specific of the nonminimally coupled theory, which depends on $\lambda$
and on the gradient of mass density, and has an effect in the ocean because of compressibility of seawater.

These results can be achieved after a suitable treatment of the conversion of pressure to depth in the ocean by resorting to the equation of state of seawater
and taking into account the effect of the extra force on hydrostatic equilibrium.
If the sole Yukawa interaction were present the experiment would yield only a bound on $\alpha$, while, in the presence of the extra force
we find an upper bound on the range: $\lambda_{\rm max}= 57.4$ km. In the interval
$1 \,{\rm m}<\lambda<\lambda_{\rm max}$ the upper bound on $\alpha$ is consistent with the constraint $\alpha<0.002$ found in Ref. \cite{Zum}.
\end{abstract}

\maketitle

\section{Introduction}

In this work we show that it is possible to constrain the parameters of a nonminimally coupled (NMC) gravity model by using the results of a geophysical experiment,
performed in 1991 by Zumberge et al. \cite{Zum} to look for deviations from Newton's inverse square law in the ocean. 
In NMC gravity the Einstein-Hilbert action functional of General Relativity (GR) 
is replaced with a more general form involving two functions $f^1(R)$ and $f^2(R)$ of the
Ricci scalar curvature $R$ of space-time \cite{BBHL}. The function $f^1(R)$ has a role analogous to $f(R)$ gravity theory 
\cite{Capoz-1,Carroll,Capoz-2,DeFTs},
and the function $f^2(R)$ multiplies the matter Lagrangian density giving rise to a nonminimal coupling between geometry and matter.

NMC gravity has been applied to several astrophysical and cosmological problems such as cosmological perturbations \cite{cosmpertur},  post-inflationary reheating \cite{reheating}, possibility to account for dark matter \cite{drkmattgal,drkmattclus}  and the current accelerated expansion of the Universe \cite{curraccel}.
The Solar System constraints were examined in Ref. \cite{SolSystConst}.
For other implications of the NMC gravity theories see Refs. \cite{puetzobuk-1,obukpuetz,puetzobuk-2,puetzobuk-3}.

In Ref. \cite{MPBD} a nonminimally coupled curvature-matter gravity model has been considered where the functions $f^1(R)$ and $f^2(R)$
have been assumed analytic at $R=0$, and the coefficients of their Taylor expansions around $R=0$ have been considered as the parameters
of the model. The metric around a spherical body with uniform mass density has been shown to be a perturbation of the weak-field Schwarzschild metric, 
particularly the perturbation of the $00$ component of the metric tensor contains a Yukawa potential. 
It has been shown that, in the nonrelativistic limit, the range $\lambda$ of the Yukawa perturbation is given by $\lambda=\sqrt{6a_2}$,
where the parameter $a_2$ of the model is proportional to the coefficient of $R^2$ in the Taylor expansion of the function $f^1(R)$.
The strength $\alpha_0$ of the Yukawa potential is given by $\alpha_0=(1-\theta)\slash 3$, where $\theta$ is the ratio $q_1\slash a_2$ and the
parameter $q_1$ is the coefficient of $R$ in the Taylor expansion of $f^2(R)$ \cite{MPBD,CPM}.
Since $f^2(R)$ multiplies the matter Lagrangian density in the
action functional, then the effect of the NMC vanishes in vacuum, however it affects the gravitational source, so that the NMC affects
only the strength of the Yukawa potential.

It was shown, in Ref. \cite{MPBD}, that the parameters of the NMC gravity model can be constrained through perturbations to perihelion precession by using
data from observations of Mercury's orbit. If the ratio $\theta$ is sufficiently close to $1$, $a_2\approx q_1$, then the
strength $\alpha_0$ of the Yukawa potential is small and the Yukawa range $\lambda$ can reach astronomical scales in the Solar System
satisfying the constraints resulting from data on Mercury's orbit \cite{MPBD}. Moreover, the resulting value of parameter $\gamma$ of the
parametrized post-Newtonian approximation is close to $1$ according to Solar System constraints on gravity \cite{CPM}.

In Refs. \cite{BBHL,BLP} the equations of motion of a perfect fluid in NMC gravity have been derived, showing the existence of an extra force
inside the fluid besides the Yukawa force. More precisely, the nonminimal coupling induces a nonvanishing covariant derivative of the
energy-momentum tensor. That leads to a deviation from geodesic motion and, consequently, the appearance of an extra force in the fluid.

In the present paper we compute the extra force in a perfect fluid for the NMC gravity model considered in Ref. \cite{MPBD} and show that the extra
force per unit volume is proportional to the gradient $\nabla\rho^2$ of squared mass density of the fluid, with coefficient
$G\lambda^2\theta^2$, where $G$ is the gravitational coupling. Since $\theta$ is constrained to be close to $1$ from
astronomical observations \cite{MPBD}, then a value of $\lambda$ of the order of the astronomical scale gives rise to an extra force which can lead to
a large perturbation of the hydrostatic equilibrium of a compressible fluid.

Constraints resulting from the hydrostatic equilibrium of a gravitating body, for instance the Sun, could then be used to impose an upper bound
on the strength of the extra force, hence on the Yukawa range $\lambda$. Such an effect was not taken into account in Ref. \cite{MPBD} since
orbits where computed around a body with uniform mass density, so that the extra force inside the body vanishes. A more stringent upper bound on
$\lambda$ is expected to be found from the condition of hydrostatic equilibrium of a compressible fluid on the Earth. 

The experiment devised in
Ref. \cite{Zum} to test the presence of a Yukawa force in the ocean is suitable for this purpose. This experiment was concluded in 1991 and never
repeated. Moreover, more recent lake and tower experiments provide more stringent constraints on the Yukawa force over the same distance
scales \cite{FT}. Nevertheless, the results from the ocean experiment in Ref. \cite{Zum}, based on measurements of gravitational acceleration 
along continuous profiles to depths of 5000 m in seawater, are particularly useful to constrain the NMC extra force. This is a consequence of
the role played by hydrostatic equilibrium of seawater in the experiment.

The result of the experiment in Ref. \cite{Zum} is an estimate of the Newtonian gravitational constant $G$ which also yields a constraint on the
presence of a Yukawa force. Such an estimate was achieved by measuring the gravitational acceleration $g(z)$ at varying depth, $z$, in the ocean 
by using a gravimeter in a submersible. The depth was computed from measuring sea pressure by resorting to the equation of state of
seawater and the condition of hydrostatic equilibrium. Hence, the extra force could not be directly measured since the gravimeter was not
immersed in seawater. Nevertheless, the extra force has an indirect effect on the measurement through the dependence of the pressure to
depth conversion on the hydrostatic equilibrium of seawater. In the present paper we show how an upper bound on the Yukawa range $\lambda$ 
can then be achieved at the geophysical scale by exploiting the compressibility properties of seawater. 
We point out that in the case of gravity models which predict a Yukawa force, 
but not a further force depending on the gradient of mass density of the fluid, the
constraints from this kind of experiments yield an upper bound on the strength of the Yukawa force only, but not on the range $\lambda$ \cite{FT}.

\section{The nonminimally coupled gravity model}

The action functional of nonminimally coupled gravity is of the form
\cite{BBHL}
\begin{equation}
S = \int \left[\frac{1}{2}f^1(R) + [1 + f^2(R)] \LL_m \right]\sqrt{-g} \, d^4x,
\end{equation}
where $f^i(R)$ (with $i=1,2$) are functions of the Ricci scalar curvature $R$, $\LL_m$ is the Lagrangian
density of matter, and $g$ is the metric determinant.
The standard Einstein-Hilbert action of GR is recovered by taking
\begin{equation}
f^1(R) = \frac{c^4}{8\pi G_N}R, \qquad f^2(R) = 0,
\end{equation}
where $G_N$ is Newton's gravitational constant.

The first variation of the action functional with respect to the metric $g_{\mu\nu}$ yields the field equations:
\begin{eqnarray}\label{field-eqs}
& &\left(f^1_R + 2f^2_R \LL_m \right) R_{\mu\nu} - \frac{1}{2} f^1 g_{\mu\nu}  \\
& &= \left(\nabla_\mu \nabla_\nu -g_{\mu\nu} \square \right) \left(f^1_R + 2f^2_R \LL_m \right)
+ \left(1 + f^2 \right) T_{\mu\nu}, \nonumber
\end{eqnarray}
where $f^i_R \equiv df^i\slash dR$. The trace of the field equations is given by
\begin{eqnarray}\label{trace}
& &\left( f^1_R + 2f^2_R \LL_m \right) R - 2f^1 + 3\square f^1_R +
6\square\left( f^2_R \LL_m \right)  \nonumber\\
& &= \left( 1 + f^2 \right) T,
\end{eqnarray}
where $T$ is the trace of the energy-momentum tensor $T_{\mu\nu}$.

A distinctive feature of NMC gravity is that the energy-momentum tensor of matter
is not covariantly conserved \cite{BBHL}. Indeed, applying the Bianchi identities to Eq. (\ref{field-eqs}), one finds that
\begin{equation}\label{covar-div-1}
\nabla_\mu T^{\mu\nu} = \frac{f^2_R }{ 1 + f^2} ( g^{\mu\nu} \LL_m - T^{\mu\nu} ) \nabla_\mu R.
\end{equation}
This property will play a crucial role in the nonrelativistic limit of hydrodynamics.

\subsection{Metric and energy-momentum tensors}

We use the following notation for indices of tensors:
Greek letters denote space-time indices ranging from 0 to 3,
whereas Latin letters denote spatial indices ranging from 1 to 3.
The signature of the metric tensor is $(-,+,+,+)$.

We consider the metric, $g_{\mu\nu}$, and energy-momentum, $T^{\mu\nu}$, tensors at the order
of approximation required to obtain the nonrelativistic limit of equations of motion.
At such an order the expansion of the metric tensor around the Minkowski metric in powers of $1\slash c$
is given by
\begin{eqnarray}
g_{00} &=& -1 + h_{00} + O\left(\frac{1}{c^4}\right),\label{g-00}\\
g_{ij} &=& \delta_{ij} + O\left(\frac{1}{c^2}\right),
\qquad g_{0i} = O\left(\frac{1}{c^3}\right),
\end{eqnarray}
where $h_{00}=O(1\slash c^2)$.

The components of the energy-momentum tensor to the relevant order are
(Ref. \cite{Wi}, Chapter 4.1):
\begin{eqnarray}
T^{00} &=& \rho c^2 + O\left(1\right), \label{T-00}\\
T^{0i} &=& \rho c v^i + O\left(\frac{1}{c}\right), \label{T-0i}\\
T^{ij} &=& \rho v^i v^j + p\delta_{ij} + O\left(\frac{1}{c^2}\right), \label{T-ij}
\end{eqnarray}
where matter is considered as a perfect fluid with matter density $\rho$, velocity field $v^i$, and pressure $p$.
The trace of the energy-momentum tensor is
\begin{equation}
T = -\rho c^2 + O\left(1\right).
\end{equation}
In the present paper we use $\LL_m = -\rho c^2+O(1)$ for the Lagrangian density of matter \cite{BLP}.

\subsection{Assumptions on functions $f^1(R)$ and $f^2(R)$}

In what follows we will denote $G_N$ the value of the Newtonian gravitational constant measured in the laboratory; we allow for the difference  $G\neq G_N$,
given the presence of a Yukawa interaction in the NMC model of gravity, as it will be discussed in the next section, and we set
\begin{equation}
\kappa = \frac{c^4}{16\pi G}.
\end{equation}
We assume that the functions $f^1(R)$ and $f^2(R)$ admit the following Taylor expansions around $R=0$,
which coincide with the ones used in Ref. \cite{reheating}:
\begin{eqnarray}
f^1(R) &=& 2\kappa \left(a_1 R + a_2 R^2\right) + O(R^3), \label{f(R)1-equation}\\
f^2(R) &=& q_1 R + O(R^2). \label{f(R)2-equation}
\end{eqnarray}
In the following, in order to recover GR when the function $f^1$ is linear (i.e., $a_2=0$) and $f^2=0$,
we set $a_1=1$.
Both the parameters $a_2$ and $q_1$ affect the nonrelativistic limit of the theory.

\section{nonrelativistic limit}

In this section we consider the nonrelativistic limit of the solution of the field equations
found in Ref. \cite{MPBD} and compute the equations of hydrodynamics of a perfect fluid in the
nonrelativistic limit. The solution of the field equations contains both the Newtonian and the Yukawa potentials with range and strength depending on NMC parameters $a_2$ and $q_1$.

\subsection{Field equations}\label{subs-fieldeq}

In this subsection we give the Ricci scalar $R$ at order $O\left(1\slash c^2\right)$ and the quantity $h_{00}$,
computed in Ref. \cite{MPBD}, which yield the nonrelativistic limit
of NMC gravity. The trace of the field Eqs. (\ref{trace}) at order $O\left(1\slash c^2\right)$ is given by
\begin{equation}\label{trace-2}
\left(\nabla^2-\frac{1}{6a_2}\right) \left(R - \frac{8\pi G}{c^2}\theta\rho\right) = -\frac{4\pi G}{3c^2 a_2}(1-\theta)\rho.
\end{equation}
In the following we assume $a_2>0$. The solution is of the Yukawa type \cite{MPBD}:
\begin{equation}\label{R2-solution}
R =\frac{8\pi G}{c^2} \theta\rho + \frac{(1-\theta)}{3c^2a_2}\Y,
\end{equation}
where $\Y$ denotes the Yukawa potential
\begin{equation}\label{Yukawa}
\Y=G\int\rho(t,\y)\frac{e^{-m|\x-\y|}}{|\x-\y|} d^3y,
\end{equation}
and
\begin{equation}\label{theta-lambda}
\theta=\frac{q_1}{a_2}, \qquad m^2=\frac{1}{\lambda^2} = \frac{1}{6a_2}.
\end{equation}
The range $\lambda$ of the Yukawa potential depends on the NMC parameter $a_2$ and
$\theta$ is a dimensionless quantity (see Ref. \cite{MPBD} for further details).

Expanding the $0-0$ component of the Ricci tensor $R_{\mu\nu}$ as
\begin{equation}
R_{00} = -\frac{1}{2}\nabla^2 h_{00} +O\left(\frac{1}{c^4}\right),
\end{equation}
and using the expression Eq. (\ref{T-00}) of $T^{00}=T_{00}$,
the $0-0$ component of the field Eqs. (\ref{field-eqs}), written at order $O\left(1\slash c^2\right)$, is
\begin{equation}\label{field-eqs-00-2}
\nabla^2 \left( h_{00} + 4 a_2 R - \frac{2q_1}{\kappa} \rho c^2\right) =
R - \frac{1}{\kappa}\rho c^2.
\end{equation}
The solution of this equation is \cite{MPBD}:
\begin{equation}\label{h00}
h_{00} = \frac{2}{c^2} \left[ U + \frac{1}{3}(1-\theta)\Y \right],
\end{equation}
where $U$ is the usual Newtonian potential
\begin{equation}
U = G \int \frac{\rho(t,\y)}{|\x-\y|} d^3y.
\end{equation}
Hence, the perturbation of the $0-0$ component of the Minkowski metric at order $O(1\slash c^2)$
(nonrelativistic limit) consists of the Newtonian potential plus a Yukawa potential
with range $\lambda$ and strength $\alpha_0$, given respectively by (see Ref. \cite{MPBD} for further details):
\begin{equation}\label{lambda-theta}
\lambda=\sqrt{6a_2}, \qquad \alpha_0=\frac{1}{3}(1-\theta)=\frac{1}{3}\left( 1-\frac{q_1}{a_2} \right).
\end{equation}
Then the constant $G$ describes the gravitational interaction of two masses located a distance $r$ apart,
as $r\rightarrow\infty$. Because of the presence of the Yukawa perturbation, if $\lambda$ is such that $r/\lambda\ll 1$ at laboratory
distances $r$, then $G$ is different from the value $G_N$ measured in the laboratory \cite{FT}.

\subsection{Equations of hydrodynamics}

The equations of hydrodynamics of a perfect fluid follow from the covariant divergence
of the energy-momentum tensor \cite{BBHL}, as given by Eq. (\ref{covar-div-1}) that we repeat for convinience:
\begin{equation}\label{covar-div-2}
\nabla_\mu T^{\mu\nu} = \frac{f^2_R }{ 1 + f^2} ( g^{\mu\nu} \LL_m - T^{\mu\nu} ) \nabla_\mu R.
\end{equation}
First we compute the $0$-th component of this equation. Using the components of the energy-momentum tensor
given by Eqs. (\ref{T-00})-(\ref{T-ij}), and taking into account that terms involving Christoffel symbols
give a contribution of order $O(1\slash c)$ to the $0$-th component of the covariant divergence of $T^{\mu\nu}$,
the left-hand side of Eq. (\ref{covar-div-2}) yields
\begin{equation}
\nabla_\mu T^{\mu 0} = \frac{\partial T^{\mu 0}}{\partial x^\mu} + O\left(\frac{1}{c}\right)=
c\frac{\partial\rho}{\partial t}+c\frac{\partial}{\partial x^i}(\rho v^i) + O\left(\frac{1}{c}\right).
\end{equation}
The right-hand side of Eq. (\ref{covar-div-2}) yields
\begin{equation}
\frac{f^2_R }{ 1 + f^2} ( g^{\mu 0} \LL_m - T^{\mu 0} ) \frac{\partial R}{\partial x^\mu} = O\left(\frac{1}{c}\right).
\end{equation}
Neglecting terms of order $O(1\slash c^2)$ the continuity equation then follows
in the nonrelativistic limit as usual:
\begin{equation}\label{contin-eq}
\frac{\partial\rho}{\partial t}+\frac{\partial}{\partial x^i}(\rho v^i) = 0.
\end{equation}
The NMC term on the right-hand side of Eq. (\ref{covar-div-2}) gives a distinctive contribution
to the spatial part of this equation that now we compute. The left-hand side yields
\begin{equation}
\nabla_\mu T^{\mu i} = \frac{\partial T^{\mu i}}{\partial x^\mu} + \Gamma^i_{00}T^{00} + O\left(\frac{1}{c^2}\right),
\end{equation}
where, using Eqs. (\ref{g-00}) and (\ref{h00}) for the metric tensor, the Christoffel symbol $\Gamma^i_{00}$
is given by
\begin{equation}
\Gamma^i_{00} = -\frac{1}{c^2}\left[\frac{\partial U}{\partial x^i}+
\frac{1}{3}(1-\theta)\frac{\partial\Y}{\partial x^i}\right] + O\left(\frac{1}{c^4}\right),
\end{equation}
and all other Christoffel symbols give contributions of order $O(1\slash c^4)$ to the $i$-th component of the covariant divergence of $T^{\mu\nu}$ and thus are neglected. Then, using the components of the energy-momentum tensor
given by Eqs. (\ref{T-00})-(\ref{T-ij}), for $i=1,2,3,$ we have
\begin{eqnarray}
\nabla_\mu T^{\mu i} &=& \frac{\partial}{\partial t}(\rho v^i)+\frac{\partial}{\partial x^j}(\rho v^i v^j)
-\rho\left[\frac{\partial U}{\partial x^i}+\frac{1}{3}(1-\theta)\frac{\partial\Y}{\partial x^i}\right]  \nonumber\\
&+&\frac{\partial p}{\partial x^i} + O\left(\frac{1}{c^2}\right).
\end{eqnarray}
Using now the continuity equation Eq. (\ref{contin-eq}), at order $O(1)$, we get
\begin{equation}\label{space-covar-div}
\nabla_\mu T^{\mu i} = \rho\frac{dv^i}{dt}-\rho\frac{\partial U}{\partial x^i}
-\frac{1}{3}(1-\theta)\rho\frac{\partial\Y}{\partial x^i}
+\frac{\partial p}{\partial x^i},
\end{equation}
where $d/dt=\partial/\partial t+v^i\partial/\partial x^i$ is the time derivative {\it following the fluid}.

For $i=1,2,3$, using Eqs. (\ref{f(R)1-equation}),(\ref{f(R)2-equation}) for functions $f^1,f^2$, the solution
for $R$, Eq. (\ref{R2-solution}), and formulas (\ref{lambda-theta}),
the right-hand side of Eq. (\ref{covar-div-2}) at order $O(1)$ yields
\begin{eqnarray}
& &\frac{f^2_R }{ 1 + f^2} ( g^{\mu i} \LL_m - T^{\mu i} ) \frac{\partial R}{\partial x^\mu}
= \frac{f^2_R }{ 1 + f^2} g^{ji} \LL_m \frac{\partial R}{\partial x^j}  \\
& &=-c^2 q_1 \rho\frac{\partial R}{\partial x^i}
= -\frac{1}{3}\theta(1-\theta)\rho\frac{\partial\Y}{\partial x^i}
-\frac{2}{3}\pi G \lambda^2\theta^2\frac{\partial\rho^2}{\partial x^i}.  \nonumber
\end{eqnarray}
Combining this equation with Eq. (\ref{space-covar-div}), for $i=1,2,3$, yields the equations of
NMC hydrodynamics for a perfect fluid in the nonrelativistic limit:
\begin{equation}
\rho\frac{dv^i}{dt} = \rho\frac{\partial U}{\partial x^i} - \frac{\partial p}{\partial x^i} +
\frac{1}{3}(1-\theta)^2\rho\frac{\partial\Y}{\partial x^i} -
\frac{2}{3}\pi G \lambda^2\theta^2\frac{\partial\rho^2}{\partial x^i}.
\end{equation}
We observe the presence of two additional terms in comparison with Eulerian equations of Newtonian hydrodynamics:
\begin{itemize}
\item[{\rm (i)}] a Yukawa force density with strength \\
 $\alpha(\theta)=(1-\theta)^2\slash 3$;
\item[{\rm (ii)}] an extra force density proportional to the gradient of squared mass density, $\rho^2$,
with coefficient of proportionality $(2\pi\slash 3)\,G\lambda^2\theta^2$.
\end{itemize}
The extra force density in (ii) has been extensively discussed in Ref. \cite{BBHL},
and for relativistic perfect fluids in Ref. \cite{BLP}.
Here we have derived the explicit expression of such a force density corresponding to the functions
$f^1(R),f^2(R)$ given by Eqs. (\ref{f(R)1-equation}),(\ref{f(R)2-equation}).

By equating $dv^i\slash dt$ to the centripetal acceleration on rotating Earth \cite{Griff}, we obtain the equations of hydrostatic equilibrium for seawater:
\begin{eqnarray}\label{hydro-equil}
\frac{\partial p}{\partial x^i} &=& \rho\frac{\partial U}{\partial x^i} +
\frac{\omega^2}{2}\rho\frac{\partial}{\partial x^i}(r\cos\phi)^2 +
\frac{1}{3}(1-\theta)^2\rho\frac{\partial\Y}{\partial x^i}  \nonumber\\
 &-& \frac{2}{3}\pi G \lambda^2\theta^2\frac{\partial\rho^2}{\partial x^i},
\end{eqnarray}
where $\omega=7.292115\times 10^{-5}$ rad/s
is the angular velocity of the Earth, $r$ is the distance to center of Earth, and $\phi$ is geocentric latitude.
These equations will be used in order to constrain the NMC parameters $a_2,q_1$ (equivalently, $\lambda,\theta$)
by means of the ocean experiment reported in Ref. \cite{Zum}.

\subsection{Motion of a test body in a static, spherically symmetric field}

In this subsection we discuss the implications of the nonrelativistic limit of NMC gravity for
the motion of a test body in the gravitational field of a static, spherically symmetric body.
The resulting constraints from Solar System observations will justify the need for further constraining
the NMC model of gravity by means of an ocean experiment as reported in Ref. \cite{Zum}.

The action for a point particle with mass $m$ is given by \cite{MPBD}:
\begin{equation}
S= mc\int d\tau [1+f^2(R)] \sqrt{ -g_{\mu\nu} \frac{dx^\mu}{d\tau}\frac{dx^\nu}{d\tau}},
\end{equation}
where $\tau$ is an affine parameter (which can be identified with proper time).
Variations with respect to $\delta x^\mu$ yield the equations of motion \cite{newtlimit},
\begin{equation}\label{geodesic}
\frac{d^2 x^\alpha}{d\tau^2} + \Gamma^\alpha_{\mu\nu} \frac{dx^\mu}{d\tau} \frac{dx^\nu}{d\tau} = \frac{f^2_R(R)}{ 1+f^2(R)} g^{\alpha\beta} \frac{\partial R}{\partial x^\beta},
\end{equation}
showing that the NMC gravity model leads to a deviation from geodesic motion \cite{BBHL,Sotiriou1}.

The nonrelativistic limit of the equations of motion of a test body in a static, spherically symmetric
field can be extracted from Ref. \cite{MPBD}, where the full relativistic equations of motion have been computed:
\begin{equation}\label{orbits}
\frac{d \mathbf{v}}{d t} = -GM_S\frac{\mathbf{r}}{ r^3} + (1-\theta)Y^\prime(r)\frac{\mathbf{r}}{r},
\end{equation}
where $M_S$ is the mass of the central attracting body,
$\mathbf{r}$ and $\mathbf{v}$ denote the radius vector and the velocity of the test body, respectively,
the prime denotes derivative with respect to $r=|\mathbf{r}|$, and $Y$ is the Yukawa potential
\begin{equation}
Y(r) = \frac{GM_S}{r}\left[ 1+\alpha_0 A(\lambda,R_S) \right] e^{-r/\lambda},
\end{equation}
where $R_S$ is the radius of the central body, the range $\lambda$ and the strength $\alpha_0$ are given by
Eq. (\ref{lambda-theta}),
and $A(\lambda,R_S)$ is a form factor which depends on the distribution of mass inside the central body
\cite{CPM,FT}.
If the central attracting body has uniform mass density, and $R_S \ll \lambda$, then
\begin{equation}
A(\lambda,R_S) = 1 + \frac{1}{10}\left(\frac{R_S}{\lambda}\right)^2 +
\frac{1}{280}\left(\frac{R_S}{\lambda}\right)^4 + \dots \,.
\end{equation}
The effect of deviation from geodesic motion is contained in the factor $(1-\theta)$ multiplying $Y^\prime$
in Eq. (\ref{orbits}), see Ref. \cite{MPBD} for the details.

If the unperturbed Newtonian orbit of the test body is elliptical, then the most significant effect of the
Yukawa perturbation in Eq. (\ref{orbits}) is an anomalous precession of the pericenter of the orbit,
with precession per revolution given by \cite{MPBD,FT}:
\begin{widetext}
\begin{eqnarray}\label{Yukawa-precession}
\delta\phi_P &=&  (1-\theta)^2\frac{\pi}{3}\left\{ 1 + e^2 \left[ \frac{3}{2} - \frac{L}{\lambda} + \frac{1}{8} \left( \frac{L}{\lambda}\right)^2 \right] \right\} \left[ 1 + \frac{1}{10} \left(\frac{R_S}{\lambda}\right)^2 \right] \left( \frac{L}{\lambda} \right)^2 \exp \left( -\frac{L}{\lambda}\right) \nonumber\\
&\approx& (1-\theta)^2\frac{\pi}{3}\left( \frac{L}{\lambda} \right)^2 \exp \left( -\frac{L}{\lambda}\right),
\end{eqnarray}
\end{widetext}
where $e$  and $L$ are the eccentricity and the {\it semilatus rectum} (i.e., the mean radius) of the orbit,
respectively, and the inequalities $e \ll 1$, $R_S/\lambda \ll 1$ are assumed.

If the Yukawa range $\lambda$ reaches astronomical values at Solar System scales, i.e., values of order of
either Sun-planets distances or the Earth-Moon distance, then astronomical tests of the Yukawa force, based
on observations of planetary precessions and Lunar Laser Ranging measurements \cite{Adel}, impose the constraint:
\begin{equation}
(1-\theta)^2 \ll 1 \quad\Longrightarrow\quad a_2\approx q_1,
\end{equation}
where Eq. (\ref{lambda-theta}) has been used. Hence, if the NMC parameter $q_1$ is close enough to $a_2$,
then the range $\lambda$ of the Yukawa force can reach astronomical values in the Solar System, still
evading the stringent constraints from astronomical tests on the Yukawa perturbation.

Nevertheless, if $(1-\theta)^2 \ll 1$ (hence, $\theta\approx 1$) and $\lambda$ is large in comparison with
the radius of Earth, then the extra force (ii) in Eq. (\ref{hydro-equil}) of hydrostatic equilibrium,
\begin{equation}
-\frac{2}{3}\pi G \lambda^2\theta^2\frac{\partial\rho^2}{\partial x^i},
\end{equation}
can become a significant perturbation of the hydrostatic equilibrium of a compressible fluid
($\partial\rho/\partial p \neq 0$) on Earth.
Hence, an upper bound on $\lambda$ has to follow from suitable experiments devised to test the presence
of a Yukawa force in a compressible fluid.
In the next section, by exploiting the compressibility properties of seawater \cite{seawater},
we discuss how an upper bound on the Yukawa range $\lambda$ can be imposed from
the measurement in the ocean of the Newtonian gravitational constant $G$ reported from the experiment of Ref. \cite{Zum}.

\section{Effects of NMC gravity in ocean experiments}

In Ref. \cite{Zum} the Newtonian gravitational constant has been measured in the ocean
by means of an experiment of Airy type (see Ref. \cite{FT}, Ch. 3).
Gravitational acceleration was measured down to 5000 m vertical lines using a submersible as a
platform for gravity measurements.
The experimental input consists of data for the gravitational accelerations
at various depths, $z$, below the surface of the ocean (see Section \ref{subsec-constr-ineq} for definitions), along with data for local
mass density $\rho(z)$ of seawater.
The main advantage in carrying out an experiment of Airy type
in the ocean is that mass density $\rho$ in the ocean is known with an accuracy better than
1 part in $10^4$, by resorting to the seawater equation of state available at the time of the experiment
\cite{Millero}.
The result of the measurements in Ref. \cite{Zum} constrains the strength
of a Yukawa modification to Newtonian gravity to be less than $0.002$ for scale lengths in the range
from 1 to 5000 m.

In the next subsections we model the theoretical contributions to the gravitational acceleration in seawater due to
Newtonian gravity and the Yukawa perturbation, respectively, and the contribution of the extra force to the
pressure to depth conversion.

\subsection{Contribution of Newtonian gravity}

Let us consider the contribution to the gravitational acceleration from the Newtonian part of the NMC model of gravity, with
gravitational constant $G$.
The Newtonian potential $U$ plus the centrifugal potential is referred to as the geopotential, 
and the level surface of the geopotential nearest to the mean sea level is denoted as the {\it geoid} \cite{HeisMor}.

Following Ref. \cite{Zum}, the contribution of Newtonian gravity is computed starting with a model for the mass density of the Earth,
described in Refs. \cite{StTu,StTu-2,Dah}, which is ellipsoidally layered beneath the topographic surface of the Earth, in the vicinity of the measurements.
Then the model is refined by applying corrections for the localized departures from the layered structure.
Let $P$ be a point inside the Earth, and let $Q$ be the point on the topographic surface such that the segment $\overline{PQ}$ is 
normal to the ellipsoid of constant mass density passing through $P$. The depth of $P$ is denoted by $z$ and
it is approximately given by the length of $\overline{PQ}$.

We denote by $\gamma$ the magnitude of the gravitational acceleration (Newtonian plus centrifugal) computed 
for $z\geq 0$ by means of the layered model.
The difference in $\gamma$ between $P$ and the point $Q$ at the surface ($z=0$), $\gamma(z)-\gamma(0)$, is predicted by
the ellipsoidally layered model with an accuracy of 1 part in $10^5$ or less \cite{StTu}. Such a precision was necessary when
comparing the raw gravity data of the experiment reported in Ref. \cite{Zum} with the theoretical prediction.
For the purpose of constraining NMC gravity, it is sufficient to use the spherical approximation of the gravity difference,
which corresponds to neglect the effects of the Earth's rotation \cite{StTu,StTu-2}:
\begin{equation}\label{Newton-1}
\gamma(z)-\gamma(0) \approx 2\frac{\gamma(0)}{r_s}z - 4\pi G \int_0^z \rho_\ell(z^\prime)dz^\prime,
\end{equation}
where $r_s$ is the distance of $Q$ to the center of Earth, and $\rho_\ell$ is the model layered mass density of the Earth.
The complete formulae of the ellipsoidal model are given in the  Appendix A where also terms of second order in $z/r_s$ are reported.

The magnitude of the acceleration in the field of the Earth, due to actual Newtonian gravity plus the centrifugal force, is represented by
\begin{equation}\label{Newton-overall}
g(z)=\gamma(z)+\delta g(z),
\end{equation}
where $\delta g(z)$ is a gravity disturbance which, in the case of the ocean experiment,
is caused by deviations from ellipsoidally layered mass density like, for instance,
a varying attraction of the seafloor topography, the presence of a sediment layer and
regional mass density variations in the soil beneath the ocean.
The seawater mass density $\rho_w$ did not exhibit significant lateral changes across the experimental site in the ocean \cite{Zum},
so that we can set for the actual seawater density $\rho_w=\rho_w(z)$.

The experimental site in Ref. \cite{Zum} was chosen in the Pacific ocean in order to minimize
gravity perturbations from the ocean-continent boundary (1000 km away) and from oceanic fracture zones.
Moreover, remote irregularities, such as large continental elevations and deep oceanic trenches,
have negligible effect because of the phenomenon of isostatic compensation \cite{HeisMor,Tur}.
By Archimedes law, topographic loads on the crust are buoyantly supported by similar but inverted undulations in the
shape of the crust-mantle boundary of the Earth: the attraction of this interface cancels at large enough distances
the gravity perturbation from the changing topography.
In Ref. \cite{Zum} the authors observe that the gravity disturbance due to the varying attraction of seafloor topography
is also largely canceled by isostatic compensation.

The magnitude $\gamma(0)$ on the topographic surface of the Earth, in the spherical approximation, is given by
\begin{equation}
\gamma(0)\approx\frac{GM_\oplus}{r_s^2},
\end{equation}
where $M_\oplus$ is the mass of the Earth.
In the absence of a Yukawa force ($G=G_N$) the value of $GM_\oplus$ is determined by means 
of several types of space measurements, with the dominant contribution resulting from laser ranging to the Lageos satellites \cite{Ries-ESW}.
In the presence of a Yukawa force, $GM_\oplus$ is different from the measured value due to
the effect of the Yukawa perturbation on the motion of satellites orbiting around Earth and involved in the measurements \cite{FT}.
However, if the Yukawa range $\lambda$ is much smaller than the mean distances of such satellites from Earth,
then the above difference is negligible. In this case, the value of $GM_\oplus$ determined by means of
space measurements is $GM_\oplus=398600441.5\times 10^6\,{\rm m^3 s^{-2}}$ \cite{Groten}.

Formulae up to the second order in polar flattening are reported in Appendix A, and they give
for $\gamma(0)$ the international gravity formula on the ellipsoid \cite{Moritz}.

\subsection{Contribution of Yukawa perturbation for $\lambda\ll R_\oplus$}\label{subs:Yukawa-contribution}

We compute the contribution to gravitational acceleration due to the Yukawa perturbation under
the assumption $\lambda\ll R_\oplus$, where $R_\oplus$ is the mean radius of the Earth.
The validity of such an assumption has to be verified \textit{a posteriori}.
We divide the region below the surface of the ocean into three subregions:
\begin{itemize}
\item[{\rm (i)}] seawater with mass density $\rho_w(z)$ for $0<z\leq z_w$;
\item[{\rm (ii)}] oceanic crust with mass density $\rho_c(z)$ for $z_w<z\leq z_c$;
\item[{\rm (ii)}] mantle with mass density $\rho_m(z)$ for $z_c<z$.
\end{itemize}
In Ref. \cite{Zum} the seawater density, $\rho_w$, varied from 1023.6 near the surface to
1050.5 ${\rm kg\,m^{-3}}$ at 5000 m depth.
Layer 1 of the oceanic crust is a sediment layer with mean thickness of 36 m and it has a negligible effect \cite{Zum}.
The average seafloor density in the region is 2690 ${\rm kg\,m^{-3}}$ \cite{Zum}, which we consider as the value of density
of layer 2 (typically 1.5 km thick), composed of extruded basalt affected by circulation of seawater through pores and cracks \cite{Stacey}.
Layer 3 is about 5 km thick with fewer pores and cracks, hence with a larger mass density, so that
the average density of the oceanic crust is $\overline{\rho}_c=2860$ ${\rm kg\,m^{-3}}$ \cite{Carlson}.
Eventually, the average mass density of the upper mantle ($z_c<z\lesssim 670$ km) is about $\overline{\rho}_m=3400$ ${\rm kg\,m^{-3}}$ \cite{Stacey}.

In order to correct the gravity measurements in Ref. \cite{Zum} for various effects,  the authors include in their computations
an Airy-Heiskanen model of isostatic compensation (see Ref. \cite{HeisMor}, Ch. 3) consisting of a
crust-mantle interface buried at a depth of 7000 m below the seafloor with the same density contrast. For our computations we use the values 
\begin{eqnarray}\label{density-values}
& &z_w=5000\mbox{ m}, \qquad z_c-z_w=7000\mbox{ m},  \nonumber\\
& & \overline{\rho}_c = 2860 \mbox{ ${\rm kg\,m^{-3}}$}, \qquad \overline{\rho}_m = 3400 \mbox{ ${\rm kg\,m^{-3}}$}.
\end{eqnarray}
Under the assumption $\lambda\ll R_\oplus$, we approximate the contributions to the Yukawa force due to seawater, oceanic crust and mantle,
by the field strength produced by two infinite slabs having mass density $\rho_w$ and $\rho_c$ and a half-space with density $\rho_m$, respectively.
We assume that the Yukawa force is exponentially suppressed beneath the upper mantle (an assumption that has to be verified {\it a posteriori}).
Then the Yukawa potential (\ref{Yukawa}), evaluated in cylindrical coordinates for $0\leq z \leq z_w$, is given by
\begin{widetext}
\begin{eqnarray}
\Y(z,\lambda) &=& 2\pi G \int_0^{+\infty} \rho_\ell(z^\prime)dz^\prime \int_0^{+\infty}
\frac{e^{-\sqrt{r^2+(z-z^\prime)^2}\slash\lambda}}{\sqrt{r^2+(z-z^\prime)^2}} rdr \\
&=& 2\pi G\lambda\left\{ \int_0^{z_w}\rho_w(z^\prime) e^{-|z-z^\prime|\slash\lambda}dz^\prime
+ \int_{z_w}^{z_c}\rho_c(z^\prime) e^{(z-z^\prime)\slash\lambda}dz^\prime
+ \int_{z_c}^{+\infty}\rho_m(z^\prime) e^{(z-z^\prime)\slash\lambda}dz^\prime \right\}. \nonumber
\end{eqnarray}
\end{widetext}
Denoting $\overline{\rho}_w$ the average mass density of seawater,
$\overline{\rho}_w=(1/z_w)\int_0^{z_w}\rho_w(z)dz$, and using Eq. (\ref{density-values}), we have
\begin{eqnarray}\label{Yukawa-potential}
\Y(z,\lambda) &=& 2\pi G\lambda^2 \left[ \overline{\rho}_w\left(2-e^{-z/\lambda}\right) \right.  \nonumber\\
&+& \left. (\overline{\rho}_m-\overline{\rho}_c)\left(e^{-z_w/\lambda}+e^{-z_c/\lambda}\right)e^{z/\lambda} \right]  \nonumber\\
&+&\Delta \Y(z,\lambda),
\end{eqnarray}
where $\Delta \Y(z,\lambda)$ is a correction which depends on the inhomogeneity of mass density.
In the following we set $\G(z,\lambda)=\partial\Y(z,\lambda)/\partial z$,  so that the magnitude of the gravitational acceleration due to Yukawa force is given by
\begin{equation}\label{Yukawa-acceleration}
\alpha(\theta) \G(z,\lambda), \qquad\mbox{with}\qquad \alpha(\theta)=\frac{1}{3}(1-\theta)^2.
\end{equation}
Using Eq. (\ref{hydro-equil}), and taking the derivative with
respect to $z$, we obtain the contribution of the Yukawa perturbation to the gravitational acceleration:
\begin{eqnarray}\label{Yukawa-diff}
\G(z,\lambda)&-&\G(0,\lambda) = 2\pi G\lambda
\left[ \overline{\rho}_w\left(e^{-z/\lambda}-1\right) \right.  \nonumber\\
&+& \left. (\overline{\rho}_m-\overline{\rho}_c)\left(e^{-z_w/\lambda}+e^{-z_c/\lambda}\right)
\left(e^{z/\lambda}-1\right)\right] \nonumber\\
&+&\Delta\left[\G(z,\lambda)-\G(0,\lambda)\right],
\end{eqnarray}
where $\Delta\left[\G(z,\lambda)-\G(0,\lambda)\right]$ is a correction depending on the inhomogeneity of mass density.
Since the contribution of the Yukawa perturbation has to be small,
we neglect disturbances caused by deviations from planarly layered mass density. Moreover, the experimental site in the Pacific ocean
was chosen with minimal relief (see Ref. \cite{Zum} for details).

At the surface of the ocean, the contribution of the Yukawa perturbation is given by
\begin{eqnarray}\label{Yukawa-0}
\G(0,\lambda) &=& 2\pi G\lambda \left[ \overline{\rho}_w+(\overline{\rho}_m-\overline{\rho}_c)\left(e^{-z_w/\lambda}+e^{-z_c/\lambda}\right) \right]  \nonumber\\
&+& \Delta \G(0,\lambda),
\end{eqnarray}
where $\Delta \G(0,\lambda)$ once again depends on the inhomogeneity of mass density.

\subsection{Contribution of the extra force}

In the experiment reported in Ref. \cite{Zum}, gravitational acceleration at various depths $z$ was measured by using a gravimeter
placed in a submersible. The gravimeter was able to measure the acceleration due to both the Newtonian and Yukawa force, however,
the contribution of the extra force could not be directly measured, since the gravimeter was not immersed in the seawater.
Nevertheless, the gravity measurement is indirectly influenced by the extra force for the following reason.

The experimental input requires data for the gravitational accelerations $g(z)$
at various depths $z$ below the surface of the ocean, along with data for local mass density $\rho_w(z)$ of seawater.
Hence depth $z$ has to be measured jointly with acceleration and mass density. In Ref. \cite{Zum} depths $z$
were determined from pressure, which was measured in the oceanic water by quartz pressure gauges.
Depth is then determined by resorting to the method of conversion of pressure to depth
of physical oceanography, which is based on the approximation of hydrostatic equilibrium and the seawater equation of state.
Pressure was measured with an accuracy better than 7 parts in $10^5$, which corresponds to an uncertainty of 0.35 m at 5000 m,
while the uncertainty associated with seawater mass density (which enters into the equation of state) was 0.5 m at the time of the experiment \cite{Zum}.
The root-sum-square depth uncertainty was 0.61 m.

The compressibility of seawater $(d\rho_w/dz\neq 0)$ yields a nonvanishing extra force.
Since the extra force constitutes a perturbation in the Eqs. (\ref{hydro-equil}) of hydrostatic equilibrium,
this force contributes to the conversion of pressure to depth, modifying the computed value of the depth, $z$.
In this section we now compute such a contribution.

The site of measurements in the ocean was chosen in order to minimize gravity perturbations also from
oceanic currents and fronts \cite{Zum}. Gravity and pressure measurements have been taken at depths below 500 m,
since velocity fluctuations in the upper few hundred meters at the experimental site are substantially larger than those in deep water \cite{Hildebr}.

Hydrostatic balance is the dominant balance within the vertical
(perpendicular to the ocean surface) momentum equation of seawater, as long as the vertical length
scales of motion are much smaller than the horizontal length scales \cite{Gill}.
Nevertheless, for the purposes of a precision experiment such as the one reported in Ref. \cite{Zum}, 
the use of the equation of hydrostatic equilibrium in a dynamic environment such as the ocean requires a preliminary discussion
(see also Ref. \cite{Hildebr}).
The following statements have to be considered valid only for the open ocean and deep water, which is the case of the experiment in Ref. \cite{Zum}.

Seawater pressure $p_w$ is the sum $p_w=p+p_d$ of the hydrostatic equilibrium pressure $p$ plus a perturbation
pressure $p_d$ due to dynamic effects \cite{Gill}. Equilibrium pressure is simply
denoted by $p$ since it will be the most frequently used. The contribution $p_d$ is due to perturbations among which the main ones are surface gravity waves,
internal gravity waves, geostrophic flow and tides \cite{Gill}. Instances of surface gravity waves are wind waves and a swell generated by a distant storm,
and their amplitude is small in comparison with ocean depth $z_w$. In this case, according to linear wave theory, the contribution of such waves 
to $p_d$ is exponentially damped with depth \cite{Gill},  and it is either negligible at depths $z$ below 500 m, where measurements have been taken
in the experiment in Ref. \cite{Zum}, or it can be filtered out as a noise component of the measured pressure $p_w$,
by computing the spectrum of perturbation pressure \cite{Will-Ha-Bo,Gla-Ho}. Internal waves occur due to seawater density gradients \cite{Gill},
their frequency is bounded from above by the Brunt-V{\"a}is{\"a}l{\"a} frequency (1 cycle per hour or less), and
their contribution to $p_w$ is either negligible or can also be filtered out since the time scales on which internal waves occur are of an hour or more \cite{Hildebr}.

Geostrophic flow is the result of the balance between the Coriolis acceleration and the horizontal pressure gradient \cite{Gill}, and it gives rise to both
a stationary contribution $H_g$ to the height of the topographic surface of the ocean above the geoid, and a
stationary contribution to $p_d$ given by $\rho_w(0)\gamma(0)H_g$. The height $H_g$ is on the order of a few decimeters, and the contribution
to the slope of the ocean surface is on the order of 1 m per 1000 km for a geostrophic current of 0.1 ${\rm m\,s^{-1}}$ \cite{Stewart}.

Tide-producing forces give rise to a periodic contribution to $p_d$. In the case of a lunar semidiurnal $M_2$ tide, which is the largest tidal constituent,
the period is 12.4 hours, which is one half of the lunar day, and the wavelength is half the circumference of the Earth at the latitude of the
experimental site \cite{Gill}. The periodic contribution $H_t$ of the tidal wave to the height of the topographic surface of the ocean above the geoid has
an amplitude on the order of 1 m or less at the experimental site \cite{Hildebr}, 
and the contribution to $p_d$ is given by $\rho_w(0)\gamma(0)H_t$.
By using a simple harmonic model, the vertical acceleration of seawater imparted by the tide is $a=\omega^2H_t$, where $\omega$
is the angular frequency of the tide, so that $a\approx 2\times 10^{-8}{\rm \,m\,s^{-2}}$ for a $M_2$ tide and $H_t=1$ m, which is negligible \cite{Hildebr}. 

The contribution to $p_d$ of geostrophic flow and tides, given by $\rho_w(0)\gamma(0)(H_g+H_t)$, can be
subtracted from the measured pressure $p_w$. Similarly, the effect on Newtonian gravity $g(z)$ of the displaced mass of seawater can be corrected. 
Since the contribution to the slope of the ocean surface is small, then a simple and suitable 
correction consists in the subtraction of the gravitational attraction of an infinite Bouguer plate \cite{HeisMor} given by 
$2\pi G\rho_w(0)(H_g+H_t)\approx 6\times 10^{-7}{\rm \,m\,s^{-2}}$ for $H_g=0.5$ m and $H_t=1$ m \cite{Hildebr}.
After implementing the correction we may set $H=0$ for the height of the topographic surface of the ocean above the geoid, so that we may refer depth $z$ to the geoid.
Eventually, if the perturbation pressure due to gravity waves has also been filtered out,
then hydrostatic equilibrium pressure $p=p_w-p_d$ may be used for pressure to depth conversion.

\subsubsection{Pressure to depth conversion}\label{sub-sub-sec-pressure-depth}

Gravity measurements in the experiment of Ref. \cite{Zum} have been corrected for tides and other dynamical effects 
(see Ref. \cite{Hildebr} for a discussion of the various corrections). Then, on the basis of the previous discussion,
the equations (\ref{hydro-equil}) of hydrostatic equilibrium for pressure $p$ can be used.
Since seawater mass density exhibits no significant lateral changes across the experimental site \cite{Zum},
we have $\rho_w=\rho_w(z)$.
Thus the vertical component of Eqs. (\ref{hydro-equil}) is given by
\begin{eqnarray}\label{hydro-equil-zeta}
\frac{1}{\rho_w}\,\frac{\partial p}{\partial z} &=& \frac{\partial U}{\partial z} +
\frac{\omega^2}{2}\frac{\partial}{\partial z}(r\cos\phi)^2 +
\frac{1}{3}(1-\theta)^2\frac{d\Y}{dz}  \nonumber\\
&-& \frac{4}{3}\pi G \lambda^2\theta^2\frac{d\rho_w}{dz},
\end{eqnarray}
where $r\approx r_s-z$. By integration along the vertical direction we obtain
\begin{eqnarray}\label{vertical-integration}
\int_{p_s}^p\frac{dp^\prime}{\rho_w} &=& \int_0^z g(z^\prime)dz^\prime +\alpha(\theta)\left[ \Y(z,\lambda)-\Y(0,\lambda) \right]  \nonumber\\
&-& \frac{4}{3}\pi G \lambda^2\theta^2 \left[ \rho_w(z)-\rho_w(0) \right],
\end{eqnarray}
where $p$ is pressure at depth $z$ below the surface of the ocean, and $p_s\approx 101325$ Pa is the pressure at the surface (atmospheric pressure).

Since mass density is discontinuous across the atmosphere-seawater and seawater-crust interfaces, then the derivative of density
that enters into the expression of the extra force is taken outside of the discontinuity surfaces, so that the derivative is defined everywhere except
at interfaces. It then follows that pressure is continuous across such interfaces.

In the sequel we give the main formulae, while technical details of the computations are reported in Appendix B.

Evaluation of $g$ in Eq. (\ref{vertical-integration}) yields
\begin{equation}\label{ocean-gravity}
g(z)= \gamma(0) + \beta(z)z + \delta g(z),
\end{equation}
where $\beta(z)$ is given by Eq. (\ref{Newton-1}),
\begin{equation}\label{beta(z)}
\beta(z) \approx 2\frac{\gamma(0)}{r_s} - 4\pi G \, \overline{\rho}_w(z),
\end{equation}
and $\overline{\rho}_w(z)$ denotes the average value of ${\rho}_w(z^\prime)$ over $(0,z)$.
An expression for $\beta(z)$ which accounts for the Earth's ellipticity is reported in Appendix B.

Note that the function $\beta(z)$ in Eq. $(\ref{beta(z)})$ is denoted by $\gamma(z)$ in oceanography, 
however we have denoted $\gamma$ the gravity computed by means of the layered mass density model.
The function $\beta(z)$ depends
weakly on $z$ through the mean value $\overline{\rho}_w(z)$.
Thus, following the practice used in physical oceanography, we consider $\beta$ constant and
we replace $\overline{\rho}_w(z)$ by $\overline{\rho}_w(z_w)=\overline{\rho}_w$.
If $G=G_N$ (absence of the Yukawa force), approximating $r_s\approx R_\oplus$,
then the value used in oceanography is $\beta_N=2.226\times 10^{-6}$ ${\rm s^{-2}}$.

The density $\rho_w$ of seawater is a function $\rho_w=\rho_w(S,t,p)$ of salinity $S$,
{\it in situ} temperature $t$ and pressure $p$ \cite{seawater} (for the various definitions of salinity see Ref. \cite{seawater}).
In the following we denote by $t$ the \textit{in situ} temperature, according to the notation in physical oceanography \cite{seawater}, since there will be no possibility
of confusion with the time variable.

By using the international equation of state of seawater \cite{seawater},
the integral of specific volume $1/\rho_w$ with respect to pressure, in Eq. (\ref{vertical-integration}), has the following expression:
\begin{equation}\label{integral-specif-vol}
\int_{p_s}^p\frac{dp^\prime}{\rho_w} = Q(p)+\Psi(S,t,p),
\end{equation}
where $Q(p)$ is a polynomial and $\Psi$ is a small quantity, called the dynamic height anomaly, which takes account of the deviation
of the physical state of seawater from the standard ocean (characterized by $S=35$ and $t=0\,^{\rm o}$C).
The main terms of the polynomial $Q(p)$ are the following \cite{seawater}:
\begin{eqnarray}\label{expr-Q-p}
& & Q(p) = 9.72661(p-p_s) - 2.2530 \times 10^{-5}(p-p_s)^2  \\
& & +  2.377 \times 10^{-10}(p-p_s)^3 - 1.66 \times 10^{-15}(p-p_s)^4 + \dots , \nonumber
\end{eqnarray}
where $Q$ is measured in ${\rm m^2\,s^{-2}}$ and $(p-p_s)$ is measured in decibars.

Let us consider the third term in the right-hand side of Eq. (\ref{vertical-integration}) which involves the extra force.
In Ref. \cite{Zum} depth $z$ is a derived quantity and not a measured quantity, 
while the measured quantities are electrical conductivity of seawater, temperature and pressure.
Thus, we express the contribution of the extra force in Eq. (\ref{vertical-integration}) as a function of salinity 
(which is closely related to measured conductivity), temperature and pressure:
\begin{equation}\label{integral-extra}
-\frac{4}{3}\pi G \lambda^2\theta^2 \left[ \rho_w(S,t,p)-\rho_w(S_s,t_s,p_s)\right],
\end{equation}
where, as previously, the subscript $s$ denotes surface values.

Following the method used in physical oceanography \cite{seawater,FofMill}, we solve approximately Eq. (\ref{vertical-integration})
with respect to $z$ as a function of $S,t,p$.
The result is the following formula for the pressure to depth conversion, which takes into account the effect of the extra force:
\begin{widetext}
\begin{eqnarray}\label{formula-total}
z(S,t,p) &\approx& \left[ \gamma(0) +\frac{1}{2}\beta \frac{Q(p)}{\gamma(0)} \right]^{-1} \{ Q(p)+\Psi(S,t,p) -\delta(S,t,p)
-  \alpha(\theta) \left[\Y(z_N(S,t,p),\lambda)-\Y(0,\lambda)\right]  \nonumber\\
&+& \frac{4}{3}\pi G \lambda^2\theta^2 \left[ \rho_w(S,t,p)-\rho_w(S_s,t_s,p_s)\right] \},
\end{eqnarray}
\end{widetext}
where $\delta(S,t,p)$ represents the contribution of the gravity disturbance $\delta g$, and 
\begin{equation}\label{formula-ocean-Newton}
z_N(S,t,p) \approx \frac{Q(p)+\Psi(S,t,p)}{\gamma(0) +\frac{1}{2}\beta_N\frac{Q(p)}{\gamma(0)}},
\end{equation}
is the conversion formula in the case of Newtonian gravity with $G=G_N$, hence absence of the Yukawa force.

The effect of the sole Yukawa force on the pressure to depth conversion gives a difference with respect to the Newtonian value $z_N$
of order of centimeters at 5,000 m for a value of $\alpha(\theta)$ of order $10^{-3}$, which is the order of magnitude
of the upper bound on $\alpha$ estimated in \cite{Zum}, and for all $\lambda\ll R_\oplus$. Hence we consider $z_N=z_N(S,t,p)$ 
as the depth function computed from $S,t,p$ in Ref. \cite{Zum}.

Formula (\ref{formula-total}) will be used in the next section to constrain the parameters
of the NMC gravity model by using the results of the experiment of Ref. \cite{Zum}.

\section{Constraints on the NMC gravity parameters}

In this section we find constraints on parameters $\lambda$ and $\theta$ of the NMC model of gravity by using the
result of the measurement of the Newtonian gravitational constant given in Ref. \cite{Zum}. The main purpose of this section
is to find an upper bound on the Yukawa range, $\lambda$, by exploiting the influence of the extra force on the pressure to
depth conversion.

Using Eqs. (\ref{Newton-1}) and (\ref{beta(z)}), the layered mass density model yields the following difference in Newtonian
gravitational acceleration between a point at depth $z$ in seawater and a point at the ocean surface:
\begin{equation}\label{layered-model-g-difference}
\gamma(z)-\gamma(0)=\beta z,
\end{equation}
where $z$ has to be expressed in terms of measured quantities $S,t,p$ according to the conversion formula (\ref{formula-total}).
Using Eq. (\ref{Yukawa-acceleration}) the contribution of the Yukawa force to the gravity difference is given by
\begin{equation}\label{Yukawa-G-difference}
\alpha(\theta)\left[\G(z,\lambda)-\G(0,\lambda)\right].
\end{equation}
In the case of Newtonian gravity with $G$ equal to the laboratory value $G_N$ (hence, in the absence of the Yukawa force), 
the gravity difference is $\beta_N z_N$, with $z_N$ given by Eq. (\ref{formula-ocean-Newton}).

Then we define the modeled gravity residual \cite{FT,StTu}, that is the excess of total gravity (the sum of (\ref{layered-model-g-difference}) plus
(\ref{Yukawa-G-difference})) over the Newtonian value with $G=G_N$:
\begin{equation}
\Delta g_m(z,z_N) = \beta z-\beta_N z_N+\alpha(\theta)\left[\G(z,\lambda)-\G(0,\lambda)\right],
\end{equation}
which, expressed as a function of measured quantities $S,t,p$ and parameters $\theta,\lambda$, reads as follows:
\begin{eqnarray}\label{gravity-residual}
\Delta g_m(S,t,p,\theta,\lambda) &=& \beta z(S,t,p)-\beta_N z_N(S,t,p)  \\
&+& \alpha(\theta)\left[\G(z_N(S,t,p),\lambda)-\G(0,\lambda)\right],  \nonumber
\end{eqnarray}
where, in the small quantity $\alpha(\theta)\G(z,\lambda)$, $z$ has been evaluated using $z_N(S,t,p)$ 
according to formula (\ref{formula-ocean-Newton}).
In the next section we will obtain an expression for the modeled gravity residual $\Delta g_m$.

\subsection{Evaluation of the modeled gravity residual}

In this section we express the modeled gravity residual $\Delta g_m$ as a function of measured quantities that characterize the physical state of seawater.
We use the following relation between the Newtonian gravitational constant $G$ at distances $r\gg\lambda$ (see the end of Subsection \ref{subs-fieldeq})
and the laboratory value $G_N$ (\cite{FT}, Appendix B):
\begin{equation}
G=\frac{G_N}{1+\alpha(\theta)\Phi(\lambda)},
\end{equation}
where $\Phi$ is a positive, increasing function such that $\Phi(\lambda)\ll 1$ for $\lambda\leq 1$ cm, and $\Phi(\lambda)\approx 1$ for $\lambda\geq 10$ m.
We set
\begin{equation}\label{Yukawa-potential-accel-rescal}
\Y(z,\lambda)=2\pi G \,\V(z,\lambda), \qquad \G(z,\lambda)=2\pi G \,\F(z,\lambda).
\end{equation}
We define the constant 
\begin{equation}
c_0 = \beta_N +4\pi G_N \overline{\rho}_w,
\end{equation}
and the following functions of pressure:
\begin{eqnarray}
c_1(p) &=& \gamma(0) +\frac{1}{2}\beta_N \frac{Q(p)}{\gamma(0)}, \nonumber\\
c_2(p) &=& c_1(p)+2\pi G_N \overline{\rho}_w\frac{Q(p)}{\gamma(0)}.
\end{eqnarray}
By using Eq. (\ref{formula-total}) with $\delta=0$ for $z(S,t,p)$ (we model gravity using the layered mass density model, so that $\delta g(z)=0$),
Eq. (\ref{formula-ocean-Newton}) for $z_N(S,t,p)$, and substituting in the expression (\ref{gravity-residual}) of the gravity residual, we find
\begin{widetext}
\begin{equation}\label{evaluated-gravity-residual}
\Delta g_m(S,t,p,\theta,\lambda) = \frac{2\pi G_N}{1+\alpha(\theta)\Phi(\lambda)}
\left\{ \frac{N(S,t,p,\theta,\lambda)}{D(p,\theta,\lambda)} + \alpha(\theta)
\left[ \F(z_N(S,t,p),\lambda) - \F(0,\lambda) \right] \right\}.
\end{equation}
\end{widetext}
The numerator $N(S,t,p,\theta,\lambda)$ is given by
\begin{widetext}
\begin{eqnarray}\label{residual-numerator}
N(S,t,p,\theta,\lambda) &=& A(\theta,\lambda)\left[ Q(p)+\Psi(S,t,p) \right] + c_1(p)
\left[ \beta_N+c_0\alpha(\theta)\Phi(\lambda) \right] \cdot \nonumber\\
&\cdot& \left\{ -\alpha(\theta)\left[ \V(z_N(S,t,p),\lambda)-\V(0,\lambda) \right]  +
\frac{2}{3}\lambda^2\theta^2 \left[ \rho_w(S,t,p)-\rho_w(S_s,t_s,p_s) \right]\right\}, \nonumber\\
\\
A(\theta,\lambda)&=&2\alpha(\theta)\Phi(\lambda)\left(1+\alpha(\theta)\Phi(\lambda)\right)\overline{\rho}_w \gamma(0).  \nonumber
\end{eqnarray}
\end{widetext}
The denominator $D(p,\theta,\lambda)$ is given by
\begin{equation}
D(p,\theta,\lambda) = c_1(p)\left[ c_1(p)+\alpha(\theta)\Phi(\lambda)c_2(p) \right].
\end{equation}
%

\subsection{Constraint inequalities}\label{subsec-constr-ineq}

In Ref. \cite{Zum} the gravitational acceleration was measured, together with $S,t,p$, along vertical tracks in seawater
using a gravimeter on a submersible.
Four continuous vertical gravity profiles have been measured for
depths $z_N=z_N(S,t,p)$ in the range $z_{N,1}\leq z_N \leq z_{N,2}$, with $z_{N,1}=500$ m and $z_{N,2}=4800$ m.

Since all the gravity measurements are relative, we consider gravity differences along a vertical track
between a point at depth $z_N$ and the point at depth $z_{N,1}$.
Now we introduce the observed gravity residual by means of the corrected gravity differences
\begin{equation}
g_c(z_N)-g_c(z_{N,1}),
\end{equation}
which are defined as the differences between raw gravimeter measurements corrected for an instrumental drift and for 
(see \cite{Zum} for further details)
\begin{itemize}
\item[{\rm (i)}] temporal variations: E\"otv\"os effect, Earth and ocean tides, vertical acceleration of the submersible;
\item[{\rm (ii)}] gravity disturbances $\delta g(z_N)$: isostatically compensated local seafloor topography, regional inhomogeneities
of mass density in the soil beneath the ocean.
\end{itemize}
The uncertainties in the various corrections are listed in Ref. \cite{Zum}.

Then the observed gravity residual $\Delta g_{\rm obs}(z_N)$ is defined as
the excess of the corrected gravity differences over the Newtonian values computed with $G=G_N$:
\begin{equation}\label{observ-gravity-residual}
\Delta g_{\rm obs}(z_N) = g_c(z_N)-g_c(z_{N,1}) - \beta_N(z_N-z_{N,1}).
\end{equation}
In the range $(z_{N,1},z_{N,2})$ the fit to the average of the slopes of $\Delta g_{\rm obs}$ in the four profiles is \cite{Zum}
\begin{equation}\label{reduced-gravity-gradient}
\frac{d\Delta g_{\rm obs}}{d z_N}=-0.060\pm 0.172,
\end{equation}
with slope measured in ${\rm mGal\,km^{-1}}$, where ${\rm 1\,mGal=10^{-3}\,cm\,s^{-2}}$.

Moreover, the average of four individual values of $\Delta g_{\rm obs}$ obtained on the bottom, at the average depth $z_N=5000$ m, 
differs from the average of the values of $\Delta g_{\rm obs}$ measured in the water column by less than 0.05 mGal \cite{Zum}.
Using Eq. (\ref{reduced-gravity-gradient}) it follows
\begin{eqnarray}\label{measured-gravity-residual}
-0.232\times 10^{-8} (z_N-z_{N,1}) &\leq& \Delta g_{\rm obs}(z_N)  \\
&\leq& 0.112\times 10^{-8} (z_N-z_{N,1}),  \nonumber
\end{eqnarray}
with the gradient of the gravity residual measured in ${\rm s^{-2}}$.

The expression for the modeled gravity residual $\Delta g_m$
requires the knowledge of seawater density $\rho_w$ and dynamic height anomaly $\Psi$ as functions
of measured quantities $S,t,p$.
The seawater density $\rho_w=\rho_w(S,t,p)$ was computed in Ref. \cite{Zum}
along the vertical profiles by using the equation of state of seawater available at the time of the experiment \cite{Millero}.
In Ref. \cite{Zum}, computed values of density $\rho_w$ are reported for the values of depth $z_N\approx 0$ at the ocean surface
and $z_N=z_w=5000\,{\rm m}$ at the bottom.

By using the data available at Ref. \cite{Zum}, in order to obtain a constraint on NMC gravity parameters $\theta,\lambda$,
we have two possibilities:
\begin{itemize}
\item[{\rm (i)}] we compute seawater density for the standard ocean, hence we set $\rho_w=\rho_w(35,0,p)$, by using the equation of state
of seawater, which corresponds to neglect $\Psi(S,t,p)$;
\item[{\rm (ii)}] we use the values of $\rho_w$ reported in Ref. \cite{Zum} for $z_N=0$ and $z_N=z_w$, computed
by using the full equation of state of seawater with $\Psi(S,t,p)$, and we apply the estimate (\ref{reduced-gravity-gradient}) of the gradient of the observed gravity residual to the range of depths $(0,z_w)$.
\end{itemize}
Here we adopt the approach (ii), but see also the discussion in the sequel. 
Then,  expanding the interval of depths from $(z_{N-1},z_N)$ to $(0,z_w)$ in the inequalities (\ref{measured-gravity-residual}),
and replacing the observed gravity residual $\Delta g_{\rm obs}$ with the modeled gravity residual $\Delta g_m$
expressed as the function (\ref{gravity-residual}) of measured quantities $S,t,p$ and parameters $\theta,\lambda$,
we achieve the following constraint on NMC gravity parameters:
\begin{eqnarray}\label{NMC-constraint-inequalities}
-0.232\times 10^{-8} z_w &\leq& \Delta g_m(S_w,t_w,p_w,\theta,\lambda)  \nonumber\\
&\leq& 0.112\times 10^{-8} z_w,
\end{eqnarray}
where $S_w,t_w,p_w$ are the measured values at the bottom of the ocean from which depth $z_w$ has been computed.

In order to compute $\Delta g_m(S_w,t_w,p_w,\theta,\lambda)$ we note that $S_w,t_w$ enter only in the quantities
$\Psi(S_w,t_w,p_w)$ and $\rho_w(S_w,t_w,p_w)$, while pressure $p_w$ enters separately in the expression (\ref{expr-Q-p}) for $Q(p)$.
A good starting point for pressure $p_w$ is found by inverting the quadratic approximation, given in Ref. \cite{Saunders} for the standard ocean,
to the Newtonian formula (\ref{formula-ocean-Newton}):
\begin{equation}\label{saunders-depth-to-pressure}
p_w \approx p_s+\frac{1}{2 a_2}\left\{ 1-a_1(\phi_g)-\left[\left(1-a_1(\phi_g)\right)^2-4a_2z_w\right]^{1\slash 2} \right\},
\end{equation}
where $\phi_g=35^{\rm o}13^\prime{\rm N}$ is the geographic latitude of the experimental site, and
\begin{eqnarray}
a_1(\phi_g)&=&\left(5.92+5.25\sin^2\phi_g\right)\times 10^{-3} \,\, {\rm m\,db^{-1}}, \nonumber\\
a_2&=&2.21\times 10^{-6} \,\, {\rm m\,db^{-2}},
\end{eqnarray}
where db denotes decibar. A more accurate solution would require the knowledge of the dynamic height anomaly $\Psi$ \cite{seawater},
but in the present paper we limit ourselves to the above approximation.

Using the values of seawater density reported in Ref. \cite{Zum}, and quoted in Subsection \ref{subs:Yukawa-contribution},
we have
\begin{eqnarray}\label{seawater-density-difference}
\rho_w(S_w,t_w,p_w)-\rho_w(S_s,t_s,p_s) &=& 1050.5 - 1023.6  \nonumber\\
&=& 26.9 \,\,{\rm kg\,m^{-3}}.
\end{eqnarray}
For the mean value $\overline{\rho}_w$ of density in the range of depths $(0,z_w)$ we have $\overline{\rho}_w\in(1023.6,1050.5)$,
and approximating the density profile by a linear profile, we have $\overline{\rho}_w=1037.05$ ${\rm kg\,m^{-3}}$.
The density for different values of pressure, although with no knowledge of salinity and temperature, can be computed by using
the equation of state of seawater for the standard ocean  ($S=35$ and $t=0\,^{\rm o}$C). For instance, using the equation
of state of Ref. \cite{Millero}, at pressure of 5000 decibars (corresponding to $z_N=4908.56$ m at latitude $30^{\rm o}$ \cite{FofMill}) the computed
value of density is $\rho_w(35,0,5000)=1050.68$ ${\rm kg\,m^{-3}}$ which is close to the value measured in Ref. \cite{Zum} at depth $z_w$.
Hence, a density profile, computed at large enough depths
by resorting to the standard ocean, together with the value of density measured at the ocean surface, which is available in Ref. \cite{Zum},
could still be used in set constraints on the NMC gravity
parameters at a suitable order of magnitude. In the following we exploit the values of density measured in Ref. \cite{Zum} which are
suitable for our purposes.

The dynamic height anomaly $\Psi$ is a small quantity that generally increases with pressure and gives a contribution to $z_N(S,t,p)$ which varies
regionally from 0 to 4 m at a pressure of 5000 decibars \cite{seawater,Saunders}. A climatological correction should be employed in order to
estimate $\Psi$ accurately. For our purposes, using Eq. (\ref{formula-ocean-Newton}), we let $\Psi(S_w,t_w,p_w)$ vary in the range
\begin{equation}
0 \leq \Psi(S_w,t_w,p_w) \leq 4\left[  \gamma(0)+\frac{1}{2}\beta_N \frac{Q(p_w)}{\gamma(0)} \right] \, {\rm m^2s^{-2}}.
\end{equation}
Since $\Psi(S_w,t_w,p_w)\ll Q(p_w)$, and $(Q+\Psi)$ is multiplied in (\ref{residual-numerator}) by a factor proportional to $\alpha(\theta)$,
then the impact of $\Psi$ on the constraints on NMC gravity parameters turns out to be negligible.

For the evaluation of the Yukawa terms $\V$ and $\F$ in the expression (\ref{evaluated-gravity-residual}) of the modeled gravity residual,
we neglect the terms $\Delta\Y$ and $\Delta\G$ depending on inhomogeneity of mass density in
Eqs. (\ref{Yukawa-potential}) and (\ref{Yukawa-diff}). Then, using Eqs. (\ref{Yukawa-potential-accel-rescal}), (\ref{Yukawa-potential}) 
and (\ref{Yukawa-diff}), we have
\begin{widetext}
\begin{eqnarray}\label{residual-Yukawa-terms}
\V(z_w,\lambda)-\V(0,\lambda)&=&\lambda^2 \left[ \overline{\rho}_w\left(1-e^{-z_w/\lambda}\right)
-(\overline{\rho}_m-\overline{\rho}_c)\left(1-e^{z_w/\lambda}\right)\left(e^{-z_w/\lambda}+e^{-z_c/\lambda}\right) \right], \nonumber\\
\F(z_w,\lambda)-\F(0,\lambda)&=&\lambda
\left[ \overline{\rho}_w\left(e^{-z_w/\lambda}-1\right)-(\overline{\rho}_m-\overline{\rho}_c)\left(1-e^{z_w/\lambda}\right)\left(e^{-z_w/\lambda}
+e^{-z_c/\lambda}\right) \right]. \nonumber\\
\end{eqnarray}
\end{widetext}
Now, substituting in the expression (\ref{evaluated-gravity-residual}) of the modeled gravity residual, a value for $\gamma(0)$
(we use the gravity formula given in Appendix A with $h=0$), 
the expression (\ref{saunders-depth-to-pressure}) of $p_w$, the expressions (\ref{residual-Yukawa-terms}) of the Yukawa terms $\V$ and $\F$,
and the values (\ref{seawater-density-difference}) of seawater density, and substituting the resulting expression of the modeled gravity residual as a function
of parameters $\theta,\lambda$ in the inequalities (\ref{NMC-constraint-inequalities}), we obtain a constraint on the NMC gravity parameters.

\begin{widetext}
	
\begin{figure}[H]
\centering
\begin{minipage}{.49\columnwidth}
	\centering
	\includegraphics[width=1.01\textwidth]{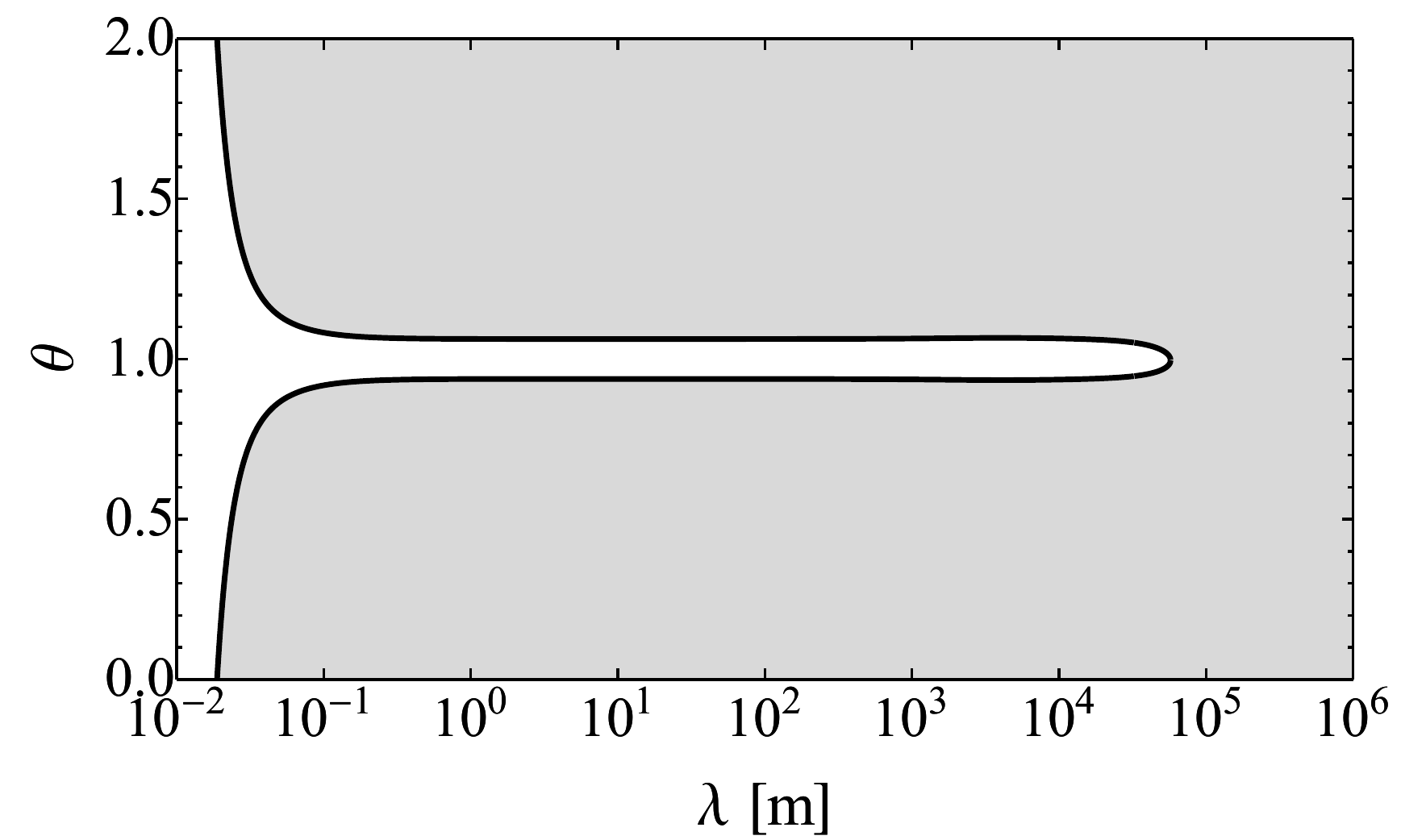}
	\caption{Constraint of Eq. (\ref{NMC-constraint-inequalities}) on the parameter plane $\lambda,\theta$.}
	\label{fig:theta_lambda}
\end{minipage}\hfill
\begin{minipage}{.49\columnwidth}
	\centering
	\includegraphics[width=1.01\textwidth]{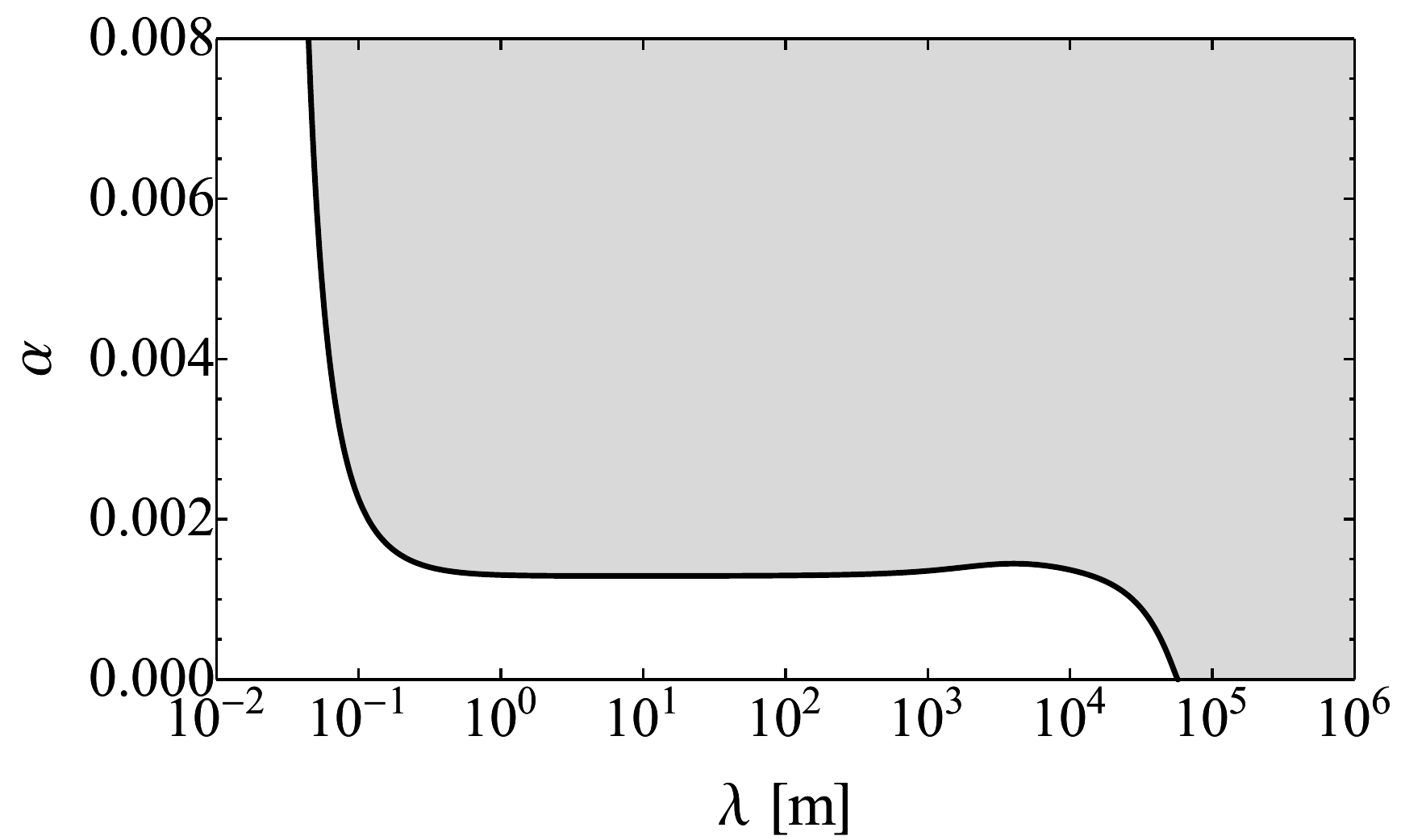}
	\caption{Values of $\alpha(\theta)$ as a function of $\lambda$ constrained by Eq. (\ref{NMC-constraint-inequalities}).}
	\label{fig:alpha_lambda}
\end{minipage}
\end{figure}
\end{widetext}

\begin{figure}[H]
 \centering
 \includegraphics[width=0.42\textwidth]{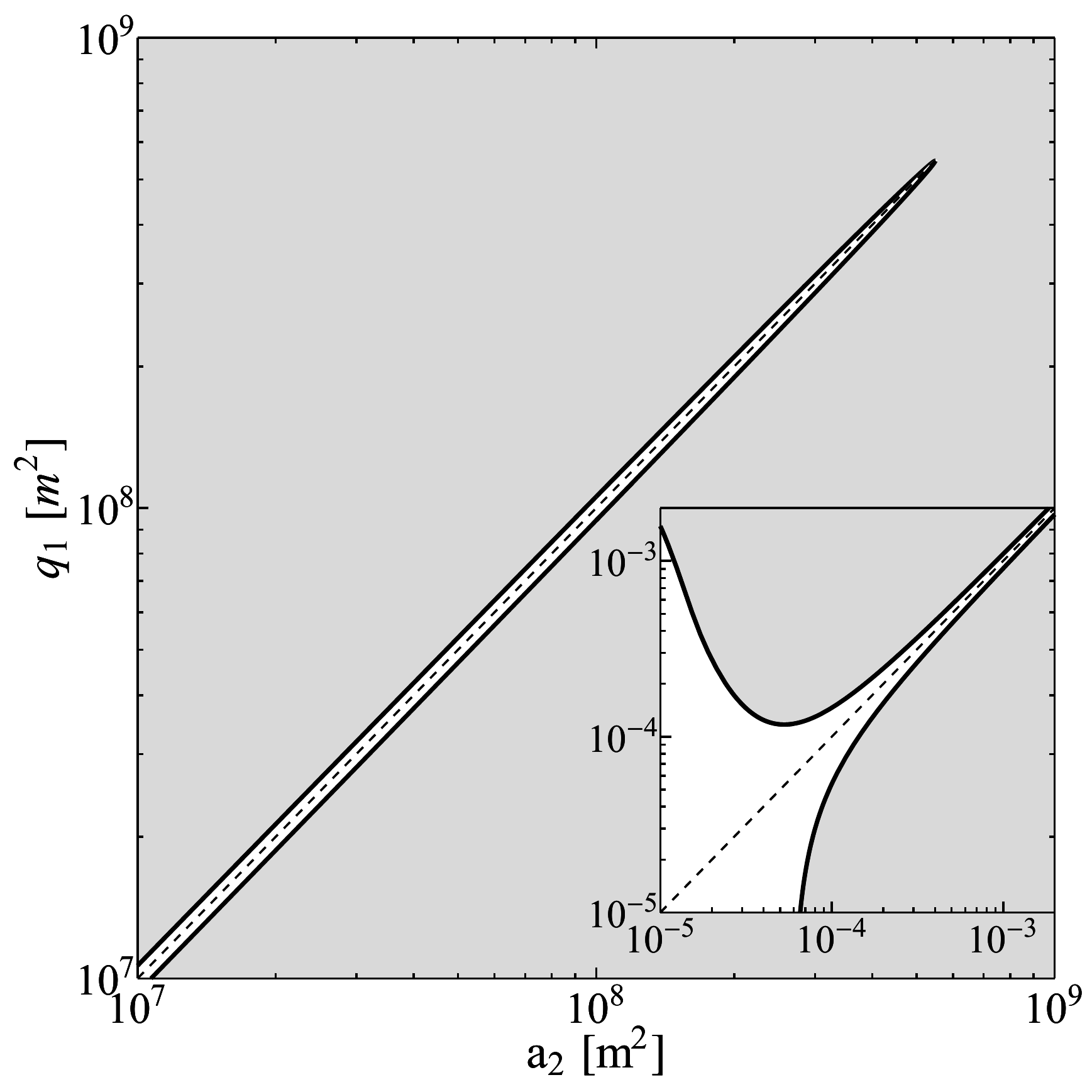}
\caption{Values of $q_1$ as a function of $a_2$ constrained by Eq. (\ref{NMC-constraint-inequalities}).}
\label{fig:q1_a2}
\end{figure}

Our results are graphically reported in Figures \ref{fig:theta_lambda}, \ref{fig:alpha_lambda} and \ref{fig:q1_a2},
in the case $\Psi(S_w,t_w,p_w)=0$. The admissible regions for the parameters of the NMC gravity model are plotted in white, while the excluded
regions are plotted in grey. Fig. \ref{fig:theta_lambda} shows the admissible region in the plane of parameters with coordinates $(\lambda,\theta)$,
Fig. \ref{fig:alpha_lambda} in the plane $(\lambda,\alpha)$, and Fig. \ref{fig:q1_a2} in the plane $(a_2,q_1)$ (we remind that
$\theta=q_1\slash a_2$ and $\lambda^2=6\,a_2$).

The plots clearly show an upper bound on the Yukawa range $\lambda$ which is located at 
$\lambda_{\rm max}=57.4$ km, so that the condition $\lambda\ll R_\oplus$ is satisfied for $\lambda\leq\lambda_{\rm max}$.
The existence of such an upper bound is a consequence of the presence of the extra force and it
is missing in the usual exclusion plots for the Yukawa perturbation where such an extra force is not considered. Fig. \ref{fig:alpha_lambda} shows that
in the range $1 \,{\rm m}<\lambda<10^4 \,{\rm m}$ the upper bound on the strength $\alpha$ of the Yukawa force is consistent with the constraint
$\alpha<0.002$ found in Ref. \cite{Zum}.

We have found that, in the range $1 \,{\rm m}<\lambda<\lambda_{\rm max}$, the contribution from the Yukawa force to pressure to depth
conversion is less than 1 cm, and the contribution from the extra force is increasing and less than 2.51 m. This last upper bound is smaller than
depth uncertainty reported in Ref. \cite{deMous} for a multibeam echo-sounder, which turned out to be between 0.1\% and 0.2\% of mean water
depth, corresponding to 5-10 m at a depth of 5000 m.

Eventually, all these results illustrate that the ocean experiment of Ref. \cite{Zum} can yield interesting results for the nonrelativistic limit 
of the nonminimally coupled curvature-matter theory of gravity.

\subsection{Relation with astronomical tests}

In this section we discuss the relation between the constraints on NMC gravity parameters obtained by means of the ocean experiment
and constraints resulting from astronomical data, particularly from the observation of Mercury's perihelion precession and Lunar geodetic precession.
In the sequel constraints from astronomical data are achieved by requiring that the Yukawa precession rate (\ref{Yukawa-precession}) is consistent
with observations of the Mercury and Moon orbits.

Recent observations of Mercury, including data from the NASA orbiter MESSENGER (Mercury Surface, Space Environment, Geochemistry and Ranging)
spacecfraft, provide a supplementary advance in Mercury perihelion \cite{Fienga-1,Fienga-2} that constrains the Yukawa force and, consequently, 
the NMC parameters $a_2,q_1$. The estimated supplementary advance (criterion 1 in the table reported in \cite{Fienga-2})
is $(0.0\pm 3.1)$ ${\rm milliarcseconds}\times{\rm cyr}^{-1}$ (${\rm mas}\times{\rm cyr}^{-1}$). 
Expressing then the precession rate (\ref{Yukawa-precession}) as a function of parameters
$\theta,\lambda$, we obtain the constraint
\begin{equation}\label{Mercury-constraint}
\delta\phi_P(\theta,\lambda) < 6.2\, T_M \,4.848 \times 10^{-11},
\end{equation}
where $T_M=0.241$ yr is the orbital period of Mercury,  the conversion milliarc seconds to radians yields a factor $4.848\times 10^{-9}$, 
and the conversion from ${\rm cy}^{-1}$ to ${\rm yr}^{-1}$ yields a further factor $10^{-2}$.
The values $L=5.546\times 10^{10}$ m, $e=0.206$ for the semilatus rectum and the eccentricity of Mercury's orbit, respectively,
and $R_S=6.957\times 10^{8}$ m for the radius of the Sun, have to be used in formula (\ref{Yukawa-precession}).

The estimate of geodetic precession of the Moon's perigee by means of Lunar Laser Ranging (LLR) is \cite{Williams}
$(1+K_{gp})\times 19.2$ ${\rm mas}\cdot{\rm yr}^{-1}$, with $K_{gp}=-0.0019\pm 0.0064$ ($K_{gp}=0$ for GR). 
This estimate yields the following LLR constraint:
\begin{equation}\label{LLR-constraint}
\delta\phi_P(\theta,\lambda) < 0.0128\cdot19.2\, T_s \,4.848 \times 10^{-9},
\end{equation}
where $T_s=0.075$ yr is the sidereal period of the Moon,
and the values $L=3.832\times 10^{8}$ m, $e=0.0549$ for the semilatus rectum and the eccentricity of the Moon's orbit, respectively,
and $R_S=R_{\oplus}=6.371\times 10^{6}$ m for the radius of the Earth, have to be used in formula (\ref{Yukawa-precession}).

\begin{figure}[H]
\centering
\includegraphics[width=0.49\textwidth]{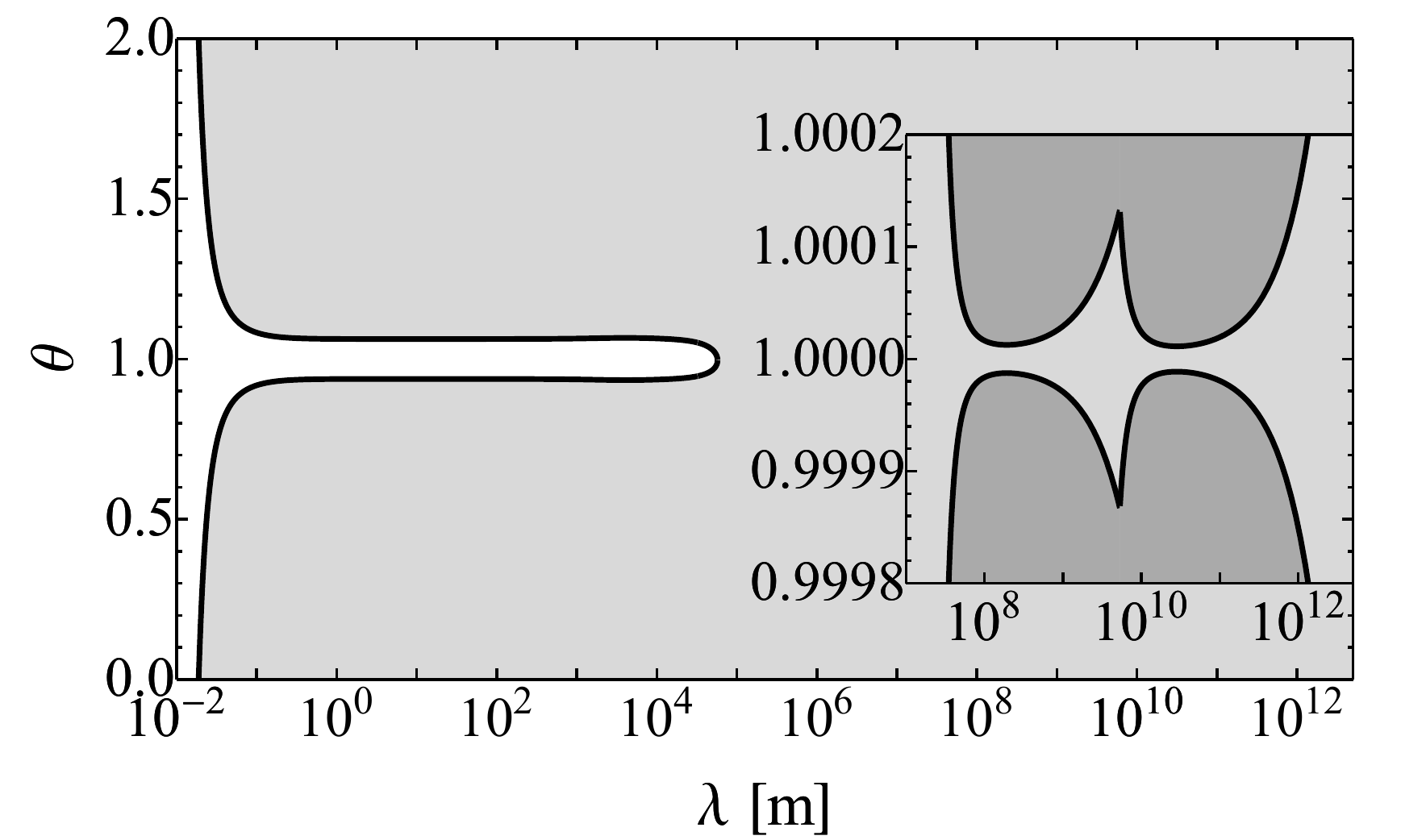}
\caption{Constraint of Eqs. (\ref{NMC-constraint-inequalities}), (\ref{Mercury-constraint}) and (\ref{LLR-constraint})
on the parameter plane $\lambda,\theta$.}
\label{fig:theta_lambda_astro}
\end{figure}

\begin{figure}[h]
 \centering
 \includegraphics[width=0.42\textwidth]{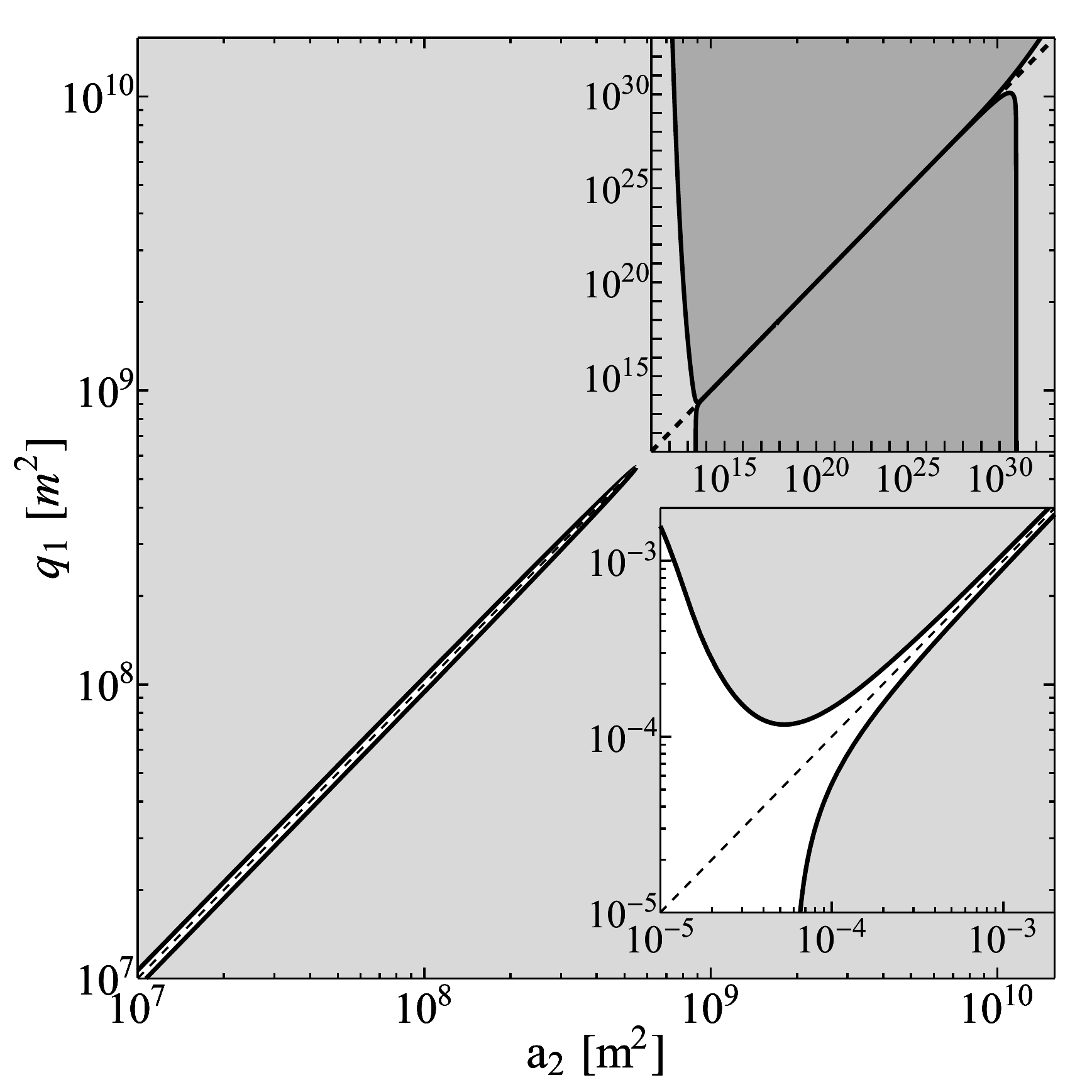}
\caption{Values of $q_1$ as a function of $a_2$ constrained by 
Eqs. (\ref{NMC-constraint-inequalities}), (\ref{Mercury-constraint}) and (\ref{LLR-constraint}).}
\label{fig:q1_a2_astro}
\end{figure}

The results are graphically reported in Figures \ref{fig:theta_lambda_astro} and \ref{fig:q1_a2_astro} which show
both the constraint from the ocean experiment and the constraints from astronomical tests. 

Figure \ref{fig:theta_lambda_astro} shows the constraints in the plane $(\lambda,\theta)$: the admissible region is plotted in white,
while the box on the right shows the constraints from LLR and Mercury using a different range of values for parameter $\theta$.
Regions inside the box which are plotted in medium grey are excluded from LLR and Mercury constraints, while regions inside the box
which are plotted in light grey are admissible for the astronomical tests, but excluded from the ocean experiment.

Figure  \ref{fig:q1_a2_astro} shows the constraints in the plane $(a_2,q_1)$: the box on the top right
shows the constraints from LLR and Mercury using a different range of values for both parameters $a_2$ and $q_1$.
The meaning of the regions plotted in medium grey and light grey inside the box is the same as in Figure  \ref{fig:theta_lambda_astro}.

Because of the upper bound on $\lambda$ at the geophysical scale from the ocean experiment, 
it turns out that excluded regions in parameter planes,
resulting from astronomical tests, are strictly contained in the excluded regions resulting from the ocean experiment.

\section{Conclusions}

In this work we have shown that the ocean experiment of Ref. \cite{Zum}, whose original purpose was searching for the deviations of the Yukawa type on the Newton's inverse square law, can be used to set up limits on the Yukawa potential arising in the nonrelativistic limit of the nonminimally coupled curvature-matter gravity theory proposed in Ref. \cite{BBHL}.
This is a rather surprising result as until this contribution the specific features of the NMC theory were believed to arise in astronomical \cite{SolSystConst,MPBD} or cosmological \cite{cosmpertur}-\cite{curraccel} contexts.

In this work we have shown that the bounds arising from Ref. \cite{Zum} are sufficiently detailed for estimating the range $\lambda$ and the strength $\alpha$
of the Yukawa potential of the nonrelativistic limit of the NMC theory.
We find an upper bound on the range, $\lambda_{\rm max}= 57.4$ km and, in the interval
$1 \,{\rm m}<\lambda<\lambda_{\rm max}$, we find an upper bound on $\alpha$ consistent with the constraint $\alpha<0.002$ found in \cite{Zum}, as it is shown
in Fig. \ref{fig:alpha_lambda}.

The upper bound on $\lambda$ is the consequence of the presence of an extra force, specific of the NMC gravity model, which 
depends itself on $\lambda$ and has an effect in an
environment  with a gradient of mass density, like seawater in the ocean. Thus the experiment of Ref. \cite{Zum} allows us to obtain an upper bound on $\lambda$
at the geophysical scale.
For sure, improvements can be expected both on the experimental and on the theoretical fronts.

Experiments inspired by the one of Ref. \cite{Zum} can be repeated and considered in other contexts. On the more theoretical side, we can hope for further constraints on the functions $f^1(R)$ and $f^2(R)$ arising from astrophysical and cosmological arguments so that more specific forms of them can be studied in the nonrelativistic limit of these gravity models.
\newpage
\appendix
\section*{Appendix A}

In this appendix we give the formulae of the contribution to the gravitational acceleration on Earth 
from Newtonian gravity plus the centrifugal force. The ellipsoidally layered model, which takes into account
the effects of Earth's rotation, yields the following expression of the gravity difference \cite{StTu,StTu-2}:
\begin{equation}
\gamma(z)-\gamma(0) = V(z) - 4\pi G X(z),
\end{equation}
where
\begin{eqnarray}
V(z) &=& 2\frac{\gamma(0)}{r_s}z \left[ 1+\frac{3}{2}\,\frac{z}{r_s}-
\frac{3}{2}J_2(3\sin^2\phi_s-1) \right]  \nonumber\\
&+& 3\omega^2 z \cos^2\phi_s,
\end{eqnarray}
and
\begin{eqnarray}
X(z) &=&  \frac{d}{a}\left[ 1+2\frac{z}{r_s}
+\frac{1}{2}\left(1-\frac{d^2}{a^2}\right) \right] \int_0^z \rho_\ell(z^\prime)dz^\prime  \nonumber\\
&-&\frac{2}{r_s}\int_0^z \rho_\ell(z^\prime)z^\prime dz^\prime.
\end{eqnarray}
In the above formulae, $r_s$ is the distance of $Q$ to the center of Earth, $\phi_s$ is the geocentric latitude of $Q$
(subscripts $s$ denote surface values),
$J_2=0.001082635$ is the quadrupole moment of the Earth, $\omega$ is the angular velocity of the Earth,
$a$ and $d$ are the semi-major and semi-minor axes of a reference ellipsoid which globally approximates the geoid \cite{HeisMor}, 
$a=6378137$ m and $(1-d^2\slash a^2)=0.0066944$, and $\rho_\ell$ is the model layered mass density of the Earth.

Neglecting the terms with $J_2$ and $\omega^2$, which depend on Earth's rotation, and neglecting terms of second order in $z/r_s$,
we get the approximation (\ref{Newton-1}), which is sufficient for the purpose of constraining NMC gravity, in the sense that the
further corrections here reported have a very small impact (not visible in the exclusion plots) on the constraints.

The distance of $Q$ from the center of Earth, to first order in polar flattening, is given by
\begin{equation}\label{distance-to-center}
r_s=a\left[1-\frac{1}{2}\left(\frac{a^2}{d^2}-1\right)\sin^2\phi_s\right]+h,
\end{equation}
where $h$ is the height of $Q$ above the reference ellipsoid.
In the following we also need the geographic latitude $\phi_g$, defined as follows:
\begin{equation}\label{geogr-latitude}
\tan\phi_g = \frac{a^2}{d^2}\tan\phi_s, \qquad \sin^2\phi_s \approx \sin^2\phi_g - \frac{a-d}{a}\sin^2 2\phi_g.
\end{equation}
The magnitude $\gamma(0)$ on the topographic surface of the Earth, to first order in polar flattening, is given by \cite{StTu-2}:
\begin{equation}\label{normal-gamma-0}
\gamma(0)=\frac{GM_\oplus}{r_s^2}\left[ 1-\frac{3}{2}J_2 (3\sin^2\phi_s-1) \right] - \omega^2 r_s\cos^2\phi_s,
\end{equation}
where $M_\oplus$ is the mass of the Earth.
Using the value of $GM_\oplus$ determined by means of space measurements,
the addition of second-order terms to Eq. (\ref{normal-gamma-0}) gives for $\gamma(0)$ the
international gravity formula on the ellipsoid \cite{Moritz}, plus a height correction dependent on $h$:
\begin{widetext}
\begin{equation}\label{gamma-ref-ellips-0}
\gamma(0)=978.0327\left( 1+0.0053024 \sin^2\phi_g - 0.0000058 \sin^2 2\phi_g \right) + \Delta(h) \quad {\rm cm\,s^{-2}},
\end{equation}
\end{widetext}
where the height correction $\Delta(h)$ is given by
\begin{equation}
\Delta(h)=-(0.30877 - 0.00045 \sin^2\phi_g)h + 0.000072 h^2,
\end{equation}
with $h$ measured in kilometers \cite{HeisMor}. The indicated distance of a point on the geoid from the reference ellipsoid is the geoidal undulation $N$. 
The values of $N$ are of the order of tens of meters and usually do not exceed $\pm 100$ m anywhere in the world.
The height $H$ of the topographic surface of the ocean
above the geoid is smaller and the maximum amplitude is roughly $\pm 1$ m \cite{Stewart}. If $Q$ is a point on the surface of the ocean, then we have
$h=H+N$.

Eventually, the value of geographic latitude of the experimental site in the northeast Pacific ocean reported in Ref. \cite{Zum} is 
$\phi_g=35^{\rm o}13^\prime{\rm N}$.

\appendix
\section*{Appendix B}

In this appendix we provide the details of the computations leading to the formulae reported in Subsection \ref{sub-sub-sec-pressure-depth}.

Gravity $g$ in Eq. (\ref{vertical-integration}) is given by
\begin{equation}
g(z)=\frac{\partial U}{\partial z} + \frac{\omega^2}{2}\frac{\partial}{\partial z}(r\cos\phi)^2.
\end{equation}
Since in the ocean $z\leq z_w=5000$ m, we have $z/r_s \ll 1$ and we use the following approximations of 
the terms $V(z)$ and $X(z)$ in the gravity difference given in Appendix A:
\begin{eqnarray}
V(z) &=& 2\frac{\gamma(0)}{r_s}z \left[ 1-
\frac{3}{2}J_2(3\sin^2\phi_s-1) \right] + 3\omega^2 z \cos^2\phi_s,  \nonumber\\ 
\label{V-approx} \\
X(z) &=&   (1-\eps)\int_0^z \rho_w(z^\prime)dz^\prime, \label{X-approx}
\end{eqnarray}
where
\begin{equation}
1-\eps = \frac{d}{a}\left[ 1+ \frac{1}{2}\left(1-\frac{d^2}{a^2}\right) \right].
\end{equation}
Using Eqs. (\ref{V-approx}), (\ref{X-approx}) and (\ref{Newton-overall}), we find formula (\ref{ocean-gravity}) with
\begin{eqnarray}
\beta(z) &=& 2\frac{\gamma(0)}{r_s} \left[ 1-
\frac{3}{2}J_2(3\sin^2\phi_s-1) \right] + 3\omega^2 \cos^2\phi_s  \nonumber\\
&-& 4\pi G (1-\eps)\overline{\rho}_w(z),
\end{eqnarray}
where $\overline{\rho}_w(z)$ is the average value of ${\rho}_w(z^\prime)$ over $(0,z)$, and $1-\eps=0.99998316$.

In the spherical approximation the expression of $\beta(z)$ is approximated by means of Eq. (\ref{beta(z)}),
which is sufficient for the purpose of constraining NMC gravity.

Integration of Eq. (\ref{ocean-gravity}) then yields
\begin{equation}\label{integral-g}
\int_0^z g(z^\prime)dz^\prime = \gamma(0)z +\frac{1}{2}\,\beta z^2
+\int_0^z \delta g(z^\prime)dz^\prime.
\end{equation}
Using Eq. (\ref{Yukawa-potential}), the contribution of the Yukawa potential to Eq. (\ref{vertical-integration}) is given by
\begin{widetext}
\begin{eqnarray}\label{integral-Yukawa}
\alpha(\theta)\left[ \Y(z,\lambda)-\Y(0,\lambda) \right] &=&
2\alpha(\theta)\pi G\lambda^2\left[ \overline{\rho}_w\left(1-e^{-z/\lambda}\right)
-  (\overline{\rho}_m-\overline{\rho}_c)\left(e^{-z_w/\lambda}+e^{-z_c/\lambda}\right) \left(1-e^{z/\lambda}\right)\right]  \nonumber\\
&+& \alpha(\theta) \left[\Delta\Y(z,\lambda)-\Delta\Y(0,\lambda)\right].
\end{eqnarray}
\end{widetext}
Collecting Eqs. (\ref{vertical-integration}), (\ref{integral-specif-vol}), (\ref{integral-extra}) and (\ref{integral-g}), then we have
\begin{eqnarray}\label{p-z-equation}
\left[\gamma(0) +\frac{1}{2}\beta z\right]z &=&
Q(p)+\Psi(S,t,p) -\int_0^z \delta g(z^\prime)dz^\prime  \nonumber\\
&-& \alpha(\theta)\left[ \Y(z,\lambda)-\Y(0,\lambda) \right] \\
&+& \frac{4}{3}\pi G \lambda^2\theta^2 \left[ \rho_w(S,t,p)-\rho_w(S_s,t_s,p_s)\right].  \nonumber
\end{eqnarray}
If we consider only the contribution from Newtonian gravity and we neglect the integral of the gravity disturbance $\delta g$, 
then Eq. (\ref{p-z-equation}) becomes
\begin{equation}\label{Newtonian-p-z-equation}
\left[\gamma(0) +\frac{1}{2}\beta z\right]z = Q(p)+\Psi(S,t,p).
\end{equation}
Following the method used in physical oceanography \cite{seawater}, Eq. (\ref{Newtonian-p-z-equation}) is solved using the standard quadratic
solution equation, but for $z^{-1}$. The solution $z_o$ used in oceanography is then given by
\begin{eqnarray}\label{formula-ocean}
z_o(S,t,p) &=& \frac{2\left[Q(p)+\Psi(S,t,p)\right]}{\gamma(0)+\sqrt{\gamma^2(0)+2\beta\left[Q(p)+\Psi(S,t,p)\right]}}  \nonumber\\
&\approx& \frac{Q(p)+\Psi(S,t,p)}{\gamma(0) +\frac{1}{2}\beta \frac{Q(p)}{\gamma(0)}},
\end{eqnarray}
where the square root has been expanded up to first order taking into account that $|\Psi|\ll Q$ and $\beta Q(p)\slash\gamma(0)\ll\gamma(0)$.
Note that, since the coefficient $\beta/2$ of $z^2$ in Eq. (\ref{Newtonian-p-z-equation}) is small, then
the same approximation of the square root in the solution for $z$ (not $z^{-1}$) of the quadratic equation (\ref{Newtonian-p-z-equation})
yields a solution $z_o$ independent of $\beta$, which is less accurate, since it corresponds to neglecting the variation of $\gamma(z)$ with $z$.

If $G=G_N$ (absence of the Yukawa force), then $\beta=\beta_N$ and the approximate conversion formula (\ref{formula-ocean}) becomes
Eq. (\ref{formula-ocean-Newton}).
With the further approximation $\beta_N Q(p)/\gamma(0)\approx \beta_N^\prime(p-p_s)$,
where $(p-p_s)$ is measured in decibars and $\beta_N^\prime$ has the same numerical value of $\beta_N$, but measured in ${\rm m\,s^{-2}decibar^{-1}}$,
Eq. (\ref{formula-ocean-Newton}) yields the formula for pressure to depth conversion which was used together with
the seawater equation of state available at the time of the experiment \cite{FofMill,SaunFof}. Then it is known that
the approximation of the square root plus this further approximation give an error in $z_N(S,t,p)$ of less than 10 cm at 10,000 m \cite{FofMill,SaunFof}.

Considering now also the contributions from the Yukawa force and the extra force, in Eq. (\ref{p-z-equation}),
in the small terms involving $\Y$ and $\delta g$ we replace $z$ with $z_N(S,t,p)$.
Then we solve the quadratic equation (\ref{p-z-equation}) for $z^{-1}$ and given $z_N(S,t,p)$ ($S,t,p$ being measured quantities),
we expand again the square root at first order retaining only the dominant term $Q(p)$, and
we obtain formula (\ref{formula-total}) for the pressure to depth conversion,
where the term involving the Yukawa potential $\Y$ has to be evaluated by using Eq. (\ref{integral-Yukawa}).

An \textit{a posteriori} evaluation of the upper bounds on the Yukawa force and extra force show that, for $\lambda>10$ cm,
the approximations made in the expansion of the square root give an error in $z$ of order of centimeters at 10,000 m.
Such an error turns out to be negligible for the purpose of constraining the extra force.

\section*{Acknowledgments}

The work of R.M. is partially supported, and the work of M.M. and S.DA is fully supported, by 
INFN (Istituto Nazionale di Fisica Nucleare, Italy), as part of the MoonLIGHT-2 experiment in the framework 
of the research activities of the Commissione Scientifica Nazionale n. 2 (CSN2).

We thank an anonymous referee whose suggestions have improved the presentation of the results.



\end{document}